%% file: uves_v3.tex
\documentclass[useAMS,usenatbib]{mn2e}

\usepackage{graphicx}
\usepackage{amssymb}
\usepackage{amsmath}
\usepackage{amstext}
\usepackage{multirow} 
\usepackage{lscape}

\newcommand{\elecd}{$n_{\rm e}$}
\newcommand{\elect}{$T_{\rm e}$}
\newcommand{\tf}{$t^{2}$}
\newcommand{\hb}{H$\beta$}
\newcommand{\ha}{H$\alpha$}

\newcommand{\foiii}{[O\thinspace{\sc iii}]}

\newcommand{\foi}{[O\thinspace{\sc i}]}
\newcommand{\foii}{[O\thinspace{\sc ii}]}
\newcommand{\fsii}{[S\thinspace{\sc ii}]}
\newcommand{\fsiii}{[S\thinspace{\sc iii}]}
\newcommand{\fni}{[N\thinspace{\sc i}]}
\newcommand{\fnii}{[N\thinspace{\sc ii}]}
\newcommand{\fariv}{[Ar\thinspace{\sc iv}]}
\newcommand{\fcliii}{[Cl\thinspace{\sc iii}]}

\newcommand{\fcliv}{[Cl\thinspace{\sc iv}]}
\newcommand{\fneiii}{[Ne\thinspace{\sc iii}]}
\newcommand{\ffeii}{[Fe\thinspace{\sc ii}]}
\newcommand{\ffeiii}{[Fe\thinspace{\sc iii}]}
\newcommand{\ffeiv}{[Fe\thinspace{\sc iv}]}
\newcommand{\fniqii}{[Ni\thinspace{\sc ii}]}
\newcommand{\fniqiii}{[Ni\thinspace{\sc iii}]}

\newcommand{\nitroi}{N\thinspace{\sc i}}
\newcommand{\nii}{N\thinspace{\sc ii}}
\newcommand{\niii}{N\thinspace{\sc iii}}

\newcommand{\silii}{Si\thinspace{\sc ii}}
\newcommand{\oi}{O\thinspace{\sc i}}
\newcommand{\oii}{O\thinspace{\sc ii}}

\newcommand{\cii}{C\thinspace{\sc ii}}

\newcommand{\neii}{Ne\thinspace{\sc ii}}

\newcommand{\sii}{S\thinspace{\sc ii}}

\newcommand{\niqii}{Ni\thinspace{\sc ii}}

\newcommand{\fariii}{[Ar\thinspace{\sc iii}]}

\newcommand{\hi}{H\,{\sc i}}
\newcommand{\hii}{H\thinspace{\sc ii}}
\newcommand{\di}{D\,{\sc i}}
\newcommand{\hei}{He\thinspace{\sc i}}
\newcommand{\heii}{He\thinspace{\sc ii}}

\newcommand{\ts}{\emph{$t^2$}}

\newcommand{\adfo}{${\rm ADF(O^{2+})}$}

\newcommand\ion[2]{${\rm #1^{#2}}$}           

\newcommand{\cmc}{{\rm cm$^{-3}$}}

\title[C and O abundances in star-forming galaxies]{Carbon and oxygen abundances from recombination lines in low-metallicity star-forming galaxies. Implications for chemical evolution%
			       \thanks{Based on observations collected at the European Southern 
			       Observatory, Chile, proposal number ESO 081.C-0113(A).}}
\author[C. Esteban et al.]
       {C. Esteban$^{1,2}$\thanks{E-mail: cel@iac.es}, 
        J. Garc{\'{\i}}a-Rojas$^{1,2}$, L. Carigi$^3$, M. Peimbert$^3$, F. Bresolin$^4$, 
	\newauthor 
	A. R. L{\'o}pez-S{\'a}nchez$^{5,6}$, A. Mesa-Delgado$^7$ \\
	$^1$Instituto de Astrof\'\i sica de Canarias, E-38200 La Laguna, Tenerife, Spain\\
        $^2$Departamento de Astrof\'\i sica, Universidad de La Laguna, E-38206, La Laguna, Tenerife, Spain\\
        $^3$Instituto de Astronom\'\i a, UNAM, Apdo. Postal 70-264, 04510 M\'exico D.F., Mexico\\
        $^4$Institute for Astronomy, 2680 Woodlawn Drive, Honolulu, HI 96822, USA\\
        $^5$Australian Astronomical Observatory, PO Box 915, North Ryde, NSW 1670, Australia\\
        $^6$Department of Physics and Astronomy, Macquarie University, NSW 2109, Australia\\
        $^7$Departamento de Astronom\'ia y Astrof\'isica, Facultad de F\'isica, Pontificia Universidad Cat\'olica de 
        		Chile, Av.~Vicu\~na Mackenna 4860,\\ 782-0436 Macul, Santiago, Chile\\}

\begin{document}

\date{Accepted 2014 June 12. Received 2014 June 12; in original form 2014 May 5}
\pagerange{\pageref{firstpage}--\pageref{lastpage}} \pubyear{2010}

\maketitle
\label{firstpage}

\begin{abstract}
 We present deep echelle spectrophotometry of the brightest emission-line knots of the star-forming galaxies He~2$-$10, Mkn~1271, NGC~3125, NGC~5408, POX~4, SDSS~J1253$-$0312, Tol~1457$-$262, Tol~1924$-$416 and the {\hii} region Hubble V in the 
 Local Group dwarf irregular galaxy NGC~6822. 
 The data have been taken with the Very Large Telescope Ultraviolet-Visual Echelle Spectrograph in the 
 3100--10420~{\AA} range. We determine electron densities and temperatures of the ionized gas from several emission-line 
 intensity ratios for all the objects. We derive the ionic abundances of C$^{2+}$ and/or O$^{2+}$ from faint pure recombination lines (RLs) 
 in several of the objects, permitting to derive their C/H and C/O ratios. 
 We have explored the chemical evolution at low metallicities analysing the C/O {\it vs.}~O/H, C/O {\it vs.}~N/O and C/N {\it vs.}~O/H relations for Galactic and extragalactic {\hii} regions and comparing with results for halo stars and DLAs. We find that {\hii} regions in star-forming dwarf galaxies occupy a different locus in the C/O {\it vs.}~O/H diagram than those belonging to the inner discs of spiral galaxies, indicating their different chemical evolution histories, and that the bulk of C in the most metal-poor extragalactic {\hii} regions should have the same origin than in halo stars. 
 The comparison between the C/O ratios in {\hii} regions and in stars of the Galactic thick and thin discs seems to give arguments to support the merging scenario for the origin of the Galactic thick disc. Finally, we find an apparent coupling between C and N enrichment at the usual metallicities determined for {\hii} regions and that this coupling breaks in very low-metallicity objects.
 \end{abstract}

\begin{keywords} {\hii} regions -- galaxies: abundances -- galaxies: dwarf -- galaxies: irregular -- galaxies: ISM -- galaxies: individual: He~2$-$10 -- galaxies: individual: Mkn~1271 -- galaxies: individual: NGC~3125 -- galaxies: individual: NGC~5408 -- galaxies: individual: NGC~6822 -- galaxies: individual: POX~4 -- galaxies: individual: SDSS~J1253$-$0312 -- galaxies: individual: Tol~1457$-$262 -- galaxies: individual: Tol~1924$-$416  
\end{keywords}

\section{Introduction} \label{intro}

Carbon is the most abundant heavy-element in the Universe after oxygen and has an indisputable biogenic significance. It is an important source of opacity and energy production in stars as well as a major constituent of interstellar dust and organic molecules. Despite its importance, we have very few determinations of its abundance in external  galaxies and they are based on the analysis of emission-line spectra of {\hii} regions. The most prominent spectral features of C require observations from space. The brightest collisionally excited lines (hereafter CELs) of C in ionized nebulae are C~{\sc iii}] 1909~{\AA} and [C~{\sc ii}] 2326~{\AA} lines in the UV and the determination of its line fluxes is severely affected by uncertainties in the reddening correction. On the other hand, the far-IR [C~{\sc ii}] 158~{$\mu$m} fine-structure line has the disadvantage that its emission arises predominantly in photodissociation regions, not in the ionized gas-phase. However, there are faint recombination lines (hereafter RLs) of \ion{C}{2+} in the optical that can be detected and measured 
in bright nebulae with the use of high-throughput spectrographs at large-aperture telescopes. The brightest of these RLs is C~{\sc ii} 4267~{\AA}, which has also the advantage of lying in a spectral zone free of blending with other emission lines. Some of us have been pioneer measuring the C~{\sc ii} 4267~{\AA} line in Galactic and extragalactic H\thinspace{\sc ii} regions using 
intermediate and high-spectral resolution spectroscopy \citep[e.g.][]{peimbertetal92,estebanetal02, estebanetal09, garciarojasesteban07, lopezsanchezetal07}. In particular, we have  obtained -- for the first time -- the C/H and C/O radial gradients of the ionized gas in the Milky Way \citep{estebanetal05, estebanetal13} and preliminary estimates for those gradients 
for the spiral galaxies M31, M33, M101 and NGC~2403 \citep[see][]{estebanetal14}. 

The C content in low-metallicity star-forming dwarf galaxies was first investigated by \citet{garnettetal95, garnettetal97, kobulnickyetal97} and \citet{kobulnickyskillman98} based on {\it Hubble Space Telescope} ($HST$) spectroscopy in the UV range that included measurements of the C~{\sc iii}] 1909~{\AA} line and also [C~{\sc ii}] 2326~{\AA} in some cases.  
These studies obtained data for a dozen metal-poor dwarf galaxies including I~Zw~18, NGC~5253, NGC~4861, and NGC~2366. These works 
confirmed a rather 
clear correlation between the C/O and O/H ratios with C showing an apparent ``secondary" behaviour with respect to O \citep{garnett04}. The observed C/O {\it vs.}~O/H trend has been interpreted as the time delay in the release of C by low- and intermediate-mass (LIM) stars with respect to the O production and/or metallicity-dependent yields of C in massive stars \citep{garnettetal99,henryetal00,carigi00,chiappinietal03}. 

The aim of the present paper is to detect and measure C~{\sc ii}  and O~{\sc ii} RLs in low-metallicity {\hii} regions of star-forming dwarf galaxies to explore the C content 
and the behaviour of the C/O ratio at low metallicities in combination with our previous data for {\hii} regions in discs of the Milky Way and other nearby spiral galaxies. These 
determinations of the C/O ratios would be more solid than those based on UV CELs because they are much less dependent on uncertainties in the reddenning law and the temperature structure of the nebulae.   

The structure of this paper is as follows. In \S\ref{observations} we describe the sample selection, observations and the data reduction procedure. 
In \S\ref{lines} we describe the emission line measurements and identifications as well as the reddening correction. 
In \S\ref{results} we present the physical conditions and ionic and total abundances determined for the sample objects. 
In \S\ref{discussion} we discuss the behaviour of the C/O {\it vs.}~O/H and C/O {\it vs.}~N/O relations in low-metallicity objects. 
Finally, in \S\ref{conclusions} we summarize our main conclusions. 
 
  \begin{figure*}
   \centering
   \includegraphics[scale=1]{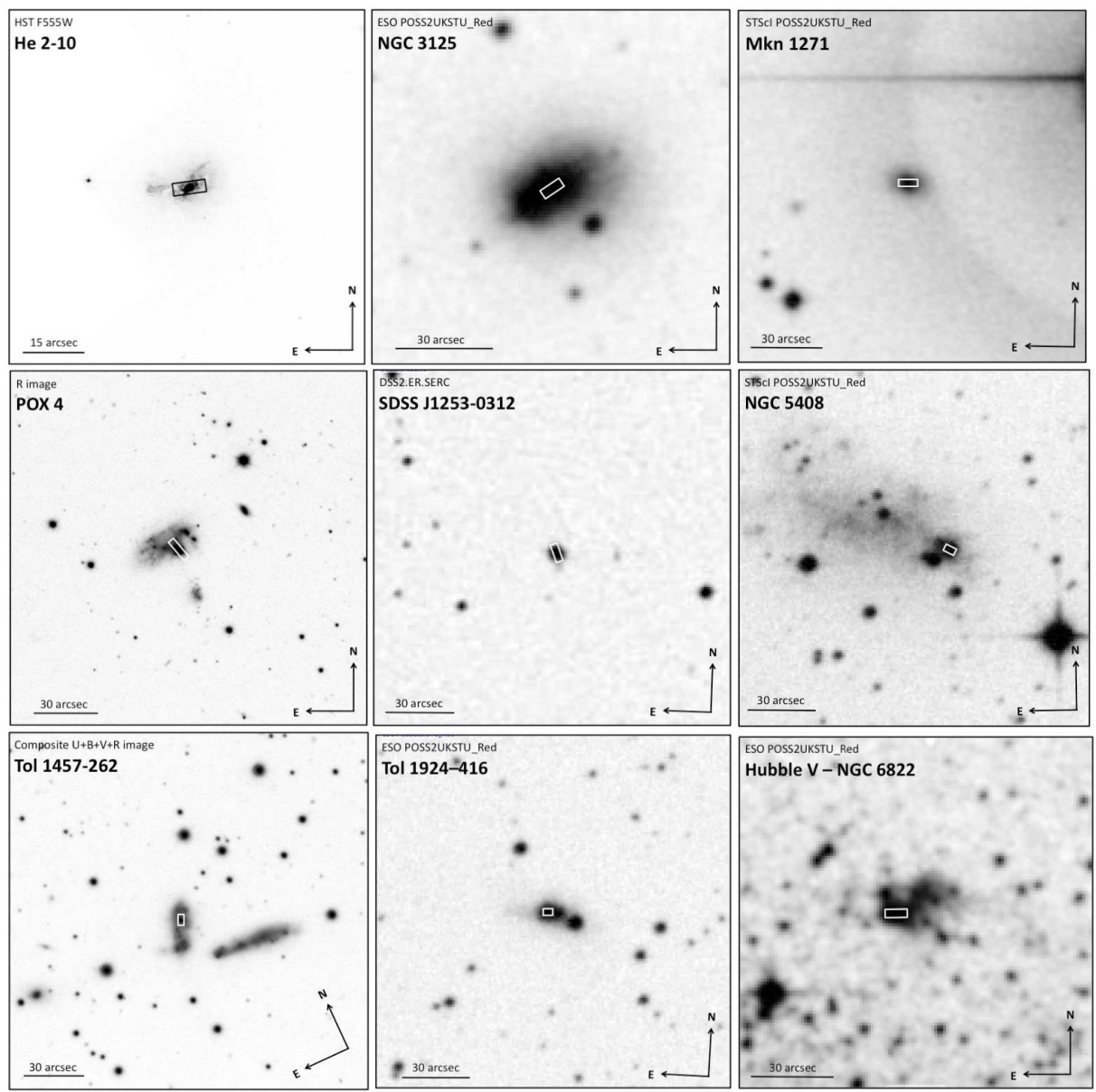} 
   \caption{Finding charts of the objects based on optical images. open rectangles indicate the position and size of the areas extracted for spectroscopical analysis. All the images have been obtained from the Aladin interactive sky atlas \citep{bonnareletal00} except those of POX~4 and Tol~1457$-$262, which have been taken from \citet{lopezsanchezesteban08}.}
   \label{charts}
  \end{figure*}

\section{Sample selection, Observations and Data Reduction} \label{observations}
 The sample was selected after an inspection of the available literature collecting spectrophotometric data of {\hii} galaxies and {\hii} regions in dwarf irregular galaxies. In order to detect and measure the faint RLs of {\cii} 4267~{\AA} and those of multiplet 1 of {\oii} at about 4650~{\AA}, we selected objects that could be observed from Paranal Observatory showing: a) high surface brightness in the H$\beta$ or H$\alpha$ lines and b) high ionization degree to ensure a better chance to detect lines of C$^{2+}$ and O$^{2+}$. We also include the bright {\hii} region Hubble~V (HV) in the dwarf irregular galaxy of the Local Group NGC~6822. This is the only object for which {\cii} and {\oii} RLs were previously detected but from low spectral resolution data \citep{apeimbertetal05}. Table~\ref{log} compiles the coordinates of the slit centre of 
our observations  as well as the morphological classification, absolute magnitudes, and distances to the galaxies as it is usually indicated in the NED database. In the case of NGC~6822, these data have been taken from the compilation of \cite{mcconnachie12} for galaxies of the Local Group. For POX~4 and Tol~1457$-$262 the values of $M_V$ have been taken from \cite{lopezsanchezesteban08}. 
In our sample objects, {\cii} and {\oii} lines were well measured --with an uncertainty better than 40\% in their 
line intensity ratio with respect to H$\beta$-- in NGC~5408, and  NGC~6822. For Mrk~1271 and NGC~3125 we obtain good measurements of {\oii} lines but only detections of {\cii}. 
For POX~4, we  measured {\cii} and only a noisy detection of the {\oii} lines. In SDSS~J1253$-$312 we detected both kinds of RLs and only {\oii} in Tol~1924$-$416. 
Unfortunately, for He 2$-$10 and Tol~1457$-$262 we did not detect the RLs of our interest. It is important to note that \cite{gusevaetal11} used the data we are analysing in this paper as part of their study 
of archive data for determining abundance patterns and the abundance discrepancy in low-metallicity emission-line galaxies. 
These authors report good measurements of both, {\cii} and {\oii} lines in NGC~5408, NGC~6822 -- like us -- but also in Mrk~1271, NGC~3125, and POX~4. On the other hand, \cite{gusevaetal11} give errors of about 35\% for the {\cii} 4267~{\AA} line in Tol~1457$-$262, and Tol~1924$-$416, objects for which we do not detect this line at all. Perhaps, these differences can be explained attending to different extraction windows used in both studies. As it is indicated ahead in this section, our 
extraction windows were restricted to isolate the brightest knots but, unfortunately, \cite{gusevaetal11} do not provide precise information about this issue.

The observations were made on 2008 May 2 and 3 at Cerro Paranal Observatory (Chile), using the UT2 (Kueyen) 
 of the Very Large Telescope (VLT) with the Ultraviolet Visual Echelle Spectrograph 
 \citep[UVES,][]{dodoricoetal00}. The standard settings of UVES were used covering the spectral 
 range from 3100 to 10420~{\AA}. Some small spectral intervals could not be observed. These are: 
 5773$-$5833, 8535$-$8649, 10081$-$10091 and 10249$-$10262~{\AA} due to the physical separation between the 
 CCDs of the detector system of the red arm and because the last orders of the spectrum do not fit completely within 
 the size of the CCD. The journal of the observations is shown in Table~\ref{log}. 
 The spectra are divided in four spectral ranges (B1, B2, R1, and R2 in Table~\ref{log}) because of the two central wavelengths used to cover the whole optical-NIR with the two arms of 
 UVES. These spectral ranges are: 3100$-$3870~{\AA} (B1), 3760$-$4985~{\AA} (B2), 4785$-$6815~{\AA} (R1) and 6700$-$10420~{\AA} (R2).
 For some of the objects, additional single short exposures of 60 seconds were taken to obtain non-saturated flux measurements for 
 the brightest emission lines. The slit was set at different position angles in the objects trying to cover the brighest areas. The 
 atmospheric dispersion corrector (ADC) was used to keep the same observed region within the slit regardless of the airmass value. 
 The slit width was set to 3$\arcsec$ ($R$ $\sim$ 15,000-20,000) as a compromise between the spectral resolution needed and the desired signal-to-noise 
 ratio of the spectra. The slit length was fixed to 10$\arcsec$ in the two bluest spectral ranges (B1 and B2) and 
 12$\arcsec$ and the two reddest ones (R1 and R2). 
 The final one-dimensional spectra we analysed were extracted for areas of different size, covering the brightest knots of the 
 objects but the same sizes in the blue and red spectral ranges. In Table~\ref{log} and Figure~\ref{charts} we indicate the position and size of the areas extracted for each object. 
 
  \begin{table*}
  \centering
   \begin{minipage}{170mm}
     \caption{Basic data of the sample galaxies and journal of observations.}
     \label{log}
    \begin{tabular}{lccccccccc}
     \hline
     & & & & & & & Extracted & \multicolumn{2}{c}{Exposure Time$^{\rm e}$}\\
     & & & & & $d^{\rm d}$ & P.A. & Area &\multicolumn{2}{c}{(s)}\\     
     Object &  R.A.$^{\rm a}$ &  Dec.$^{\rm a}$ & Type$^{\rm b}$ & $M^{\rm c}$ & [Mpc] & [$^\circ$] &  [arcsec$^2$] & B1 and R1 & B2 and R2 \\
     \hline
     He 2$-$10 & 08 36 15.0 & $-$26 24 33 & I0? pec & $-$19.0(R) & 10.5 & 98 & 8$\times$3 & 60, 3$\times$400 & 60, 3$\times$1600 \\
     NGC~3125 & 10 06 33.6 & $-$29 56 08 & BCDG & $-$19.2(R) & 12.3 & 124 & 7$\times$3 & 3$\times$400 & 60, 3$\times$1600 \\
     Mkn~1271 & 10 56 09.1 & +06 10 22 & Compact & $-$16.3(F$_{pg}$) & 13.7 & 90 & 8$\times$3 & 60, 3$\times$400 & 60, 3$\times$1600 \\
     POX~4& 11 51 11.7 & $-$20 35 58 & {\hii} & $-$19.1(V) & 47 & 40 & 8$\times$3 & 3$\times$400 & 3$\times$1600 \\
     SDSS~J1253$-$0312 & 12 53 06.0 & $-$03 12 59 & {\hii} & $-$19.8(g) & 92 & 20 & 8$\times$3 & 3$\times$400 & 3$\times$1600 \\
     NGC~5408 & 14 03 15.5 & $-$41 22 24 & IB(s)m & $-$17.2(V) & 4.9 & 62.4 & 5$\times$3 & 60, 3$\times$400 & 60, 3$\times$1800 \\
     Tol 1457$-$262 & 15 00 27.1 & $-$26 26 57 & {\hii} & $-$19.9(V) & 68 & 155 & 5.5$\times$3 & 3$\times$400 & 3$\times$1600 \\
     Tol 1924$-$416 & 19 27 58.0 & $-$41 34 27 & pec {\hii} & $-$19.9(V) & 39 & 91.7 & 4.25$\times$3 & 3$\times$400 & 3$\times$1600 \\
     NGC~6822 (HV) &  19 44 52.4 & $-$14 43 13 & IB(s)m & $-$15.2(V) & 0.46 & 91 & 8$\times$3 & 3$\times$400 & 60, 3$\times$1600 \\
     \hline
    \end{tabular}
    \begin{description}
      \item[$^{\rm a}$] Coordinates of the slit centre (J2000.0).  
      \item[$^{\rm b}$] Morphological type from NED database.
      \item[$^{\rm c}$] {Absolute magnitude from NED database except for POX~4 and Tol 1457$-$262 (L\'opez-S\'anchez \& Esteban 2008) and NGC~6822 (McConnachie 2012). The band of the photometric data is in parenthesis.}
      \item[$^{\rm d}$] Distance taken from NED database except for NGC~6822 (McConnachie 2012). 
      \item[$^{\rm e}$] Spectral ranges observed B1: 3100$-$3870 \AA; B2: 3760$-$4985 \AA; R1: 4785$-$6815 \AA; R2: 6700$-$10420~{\AA}.         
    \end{description}
   \end{minipage}
  \end{table*}

 The spectra were reduced using the {\sc iraf}\footnote{{\sc iraf} is distributed by NOAO, which 
 is operated by AURA, under cooperative agreement with NSF.} echelle reduction package, following 
 the standard procedure of bias subtraction, aperture extraction, flatfielding, wavelength 
 calibration and flux calibration. The standard stars LTT 3864, CD$-$32$^\circ$9927, and  EG~274 \citep{hamuyetal92,%
 hamuyetal94} were observed to perform the flux calibration. 
 
  \begin{figure}
   \centering
   \includegraphics[scale=0.6]{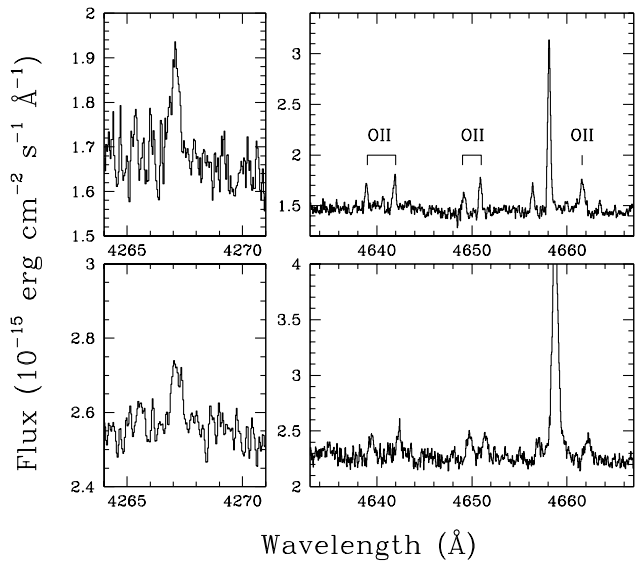} 
   \caption{Sections of the UVES spectra -- uncorrected for reddening -- of HV (NGC~6822, upper panels) and NGC~5408 (lower panels) showing the recombination 
            lines of {\cii} 4267~{\AA} (left) and multiplet 1 of {\oii} about 4650~{\AA} (right). The wavelengths have been corrected to their rest values.}
   \label{spectra}
  \end{figure}    	 

\section{Line intensities} \label{lines}

  \input{table_lines_1} 

  \input{table_lines_2} 
  
  \input{table_lines_3} 

Line fluxes were measured with the {\sc splot} routine of {\sc iraf} by integrating all the flux in the line between two given 
limits and over the average local continuum. In the case of line blending, a  
double Gaussian profile fit procedure was applied to measure the 
individual line intensities. 

 All line fluxes of a given spectrum have been normalized to a particular bright emission line 
 in each spectral range. For the bluest range (B1), the reference line was H10 3798~{\AA}. 
 For the B2 and R1 ranges, H$\beta$ was used. For the reddest 
spectral interval, R2, the reference line was {\fsii}6717~{\AA}. In order to produce a final homogeneous 
set of line flux ratios, all of them were rescaled to H$\beta$ considering the $F$(H10)/$F$(H$\beta$) and 
$F$({\fsii} 6717 \AA)/$F$(H$\beta$) line intensity ratios measured in the B2 and R1 spectral ranges, respectively. 

 The spectral ranges observed present some overlapping at the edges. The final flux of a line in the 
 overlapping regions was the average of the values obtained in both adjacent spectra. A similar procedure 
 was considered in the case of lines present in two consecutive spectral orders. The average of both 
 measurements were considered as the adopted value of the 
 line flux. In all cases, the differences in the line flux measured for the same line in 
 different orders and/or spectral ranges do not show systematic trends and are always within the 
 uncertainties.

 The identification and laboratory wavelengths of the lines were obtained following our previous works 
 on echelle spectroscopy of bright extragalactic \hii\ regions \citep{lopezsanchezetal07,estebanetal09}. 
 For a given line, the observed wavelength is determined by the central point of the two extremes of the baseline chosen for the 
line flux integration or the centroid of the line when a Gaussian fit is used (in the case of 
line blending). The final adopted values of the observed wavelength of a given line are relative 
to the heliocentric reference frame.  

The observed line intensities must be corrected for interstellar reddening. This was done using the reddening constant, 
$c$(H$\beta$), obtained from the intensities of the Balmer lines. However, the fluxes of {\hi} lines may be also 
affected by underlying stellar absorption. Consequently, we have performed an iterative procedure to derive both 
$c$(H$\beta$) and the equivalent widths of the absorption in the {\hi} lines, $W_{\rm abs}$, which we 
use to correct the observed line intensities. We assumed that the equivalent width of
the absorption components is the same for all the Balmer lines
and used the relation given by \citet{mazzarellaboroson93} to the absorption correction for each Balmer line following 
the procedure outlined by \citet{lopezsanchezetal06}. We have used the reddening curve of \cite{cardellietal89} --assuming 
 $R_V$ = 3.1-- and the observed H$\alpha$/H$\beta$, H$\gamma$/H$\beta$, H$\delta$/H$\beta$, and H$\epsilon$/H$\beta$ line ratios. We have considered the theoretical line ratios expected for case B recombination given by \citet{storeyhummer95} for electron densities of 
100 cm$^{-3}$ and a first estimation of the electron temperature of each object based on the line intensity ratios not corrected for reddening. 
At the end of tables~\ref{lines1}, \ref{lines2} and \ref{lines3}, we include the $c$(H$\beta$) and $W_{\rm abs}$ pairs that provide the 
best match between the corrected and the theoretical line ratios. In the tables we also include the observed --uncorrected for reddening-- 
integrated H$\beta$ flux, $F$(H$\beta$) and the equivalent width of this line, $W$(H$\beta$). 
 
 Tables~\ref{lines1}, ~\ref{lines2}, and ~\ref{lines3} show all the emission-line intensities measured and provide the most complete collection of emission lines available for the sample objects. For comparison, while \cite{gusevaetal11} --using the same dataset-- give line intensity ratios for 46 and 45 lines for NGC~5408 and HV (NGC~6822), respectively, we identify and 
 measure 177 and 166 lines in these two objects. Our list of identified emission lines is also more complete for the rest of the objects.  
 Columns 1 to 3 of the tables give the laboratory wavelength and 
 the identification --ion and multiplet-- of each line. Column 4 indicates the $f$($\lambda$) value of the extinction curve for 
 each line. Columns 5, 7, and 9 give the observed wavelength of the lines --corrected for heliocentric velocity-- measured in each of the 3 objects included in each table. Finally, 
 columns 6, 8, and 10 give the dereddened intensity line ratios with respect to H$\beta$ and their associated uncertainty. 
 The quoted errors include the uncertainties in line intensity measurement and error propagation in the reddening coefficient.
 
 As an example, Figure~\ref{spectra} plots sections of our flux-calibrated echelle spectra showing the recombination 
 lines of {\cii} 4267~{\AA} and multiplet 1 of {\oii} around 4650~{\AA} in the case of HV (NGC~6822) and NGC~5408.

\section{Results}\label{results}     

\subsection{Physical Conditions } \label{conditions}

  \begin{table*}
  \centering
   \begin{minipage}{180mm}
     \caption{Atomic dataset used for collisionally excited lines.}
     \label{atomic}
    \begin{tabular}{lcc}
     \hline 
	& Transition  & Collisional \\
	Ion & Probabilities & Strengths \\
     \hline 
N$^+$ & \cite{froesefischertachiev04} & \cite{tayal11} \\
O$^+$ & \cite{froesefischertachiev04} & \cite{kisieliusetal09} \\
O$^{2+}$ &  \cite{froesefischertachiev04} & \cite{palayetal12} \\
               &  					        & \cite{aggarwalkeenan99}\\
Ne$^{2+}$ & \cite{froesefischertachiev04} & \cite{McLaughlinBell00} \\
S$^+$ & \cite{Podobedovaetal09} & \cite{TayalZatsarinny10} \\
           &  \cite{TayalZatsarinny10} & \\
S$^{2+}$ &  \cite{Podobedovaetal09} & \cite{galavisetal95} \\
Ar$^{2+}$ & \cite{mendozazeippen83} & \cite{galavisetal95} \\
Fe$^{2+}$ &  \cite{quinet96} & \cite{zhang96} \\
                 & \cite{Johanssonetal00}  &  \\
     \hline
    \end{tabular}
   \end{minipage}
  \end{table*}


  \begin{table*}
   \centering
   \begin{minipage}{180mm}
   \caption{Physical Conditions.}
   \label{physcond}
    \begin{tabular}{lcccccc}
     \hline
     Parameter & Lines & HV (NGC~6822) & NGC~5408 &  Tol~1924$-$416 & NGC~3125 & Mkn~1271 \\  
     \hline
{\elecd} ({\cmc}) & {\fni} & 5700: & $-$ & $-$ & $-$ & $-$ \\
 & {\fsii} & 90$^{+110}_{-90}$ & 200 $\pm$ 110 & 90$^{+100}_{-90}$ & 110 $\pm$ 100 &  130 $\pm$ 100 \\
 & {\foii} & 100 $\pm$ 70 & 170 $\pm$ 70 & 100 $\pm$ 70 & 120 $\pm$ 80 & 160 $\pm$ 80 \\
 & {\ffeiii} & 300 - 2700 & 250 - 700 & 130 - 740 & 10 - 400 & 200 - 400 \\
 & {\fcliii} &  370$^{+530}_{-370}$ &  $-$ & 1830$^{+1960}_{-1830}$ &  210$^{+1630}_{-210}$ & $-$ \\
 & {\fariv} & 130$^{+720}_{-130}$ &  870 $\pm$ 570 & 430$^{+900}_{-430}$ & 190$^{+1230}_{-190}$ & $-$ \\
{\elect} (K) &  {\fnii} & 11800 $\pm$ 1500 & 12080  $\pm$  1100 & $-$ & $-$ & $-$ \\
  & {\foii} & 12600 $\pm$ 480 & 14000 $\pm$ 510 & 13200 $\pm$ 460 & 11600 $\pm$ 440 & 10900 $\pm$ 440 \\
 & {\fsii} & 12300 $\pm$ 700 & 14100 $\pm$ 820 & 14400 $\pm$ 1000 & 11850 $\pm$ 750 & 10030 $\pm$ 670 \\
& {\bf $T$(low)} & {\bf 12470 $\pm$ 440} & {\bf 13850 $\pm$ 720} & {\bf 13430 $\pm$ 620} & {\bf 11650 $\pm$ 400} & {\bf 10670 $\pm$ 530} \\
 & {\foiii} & 11050 $\pm$ 240 & 15300 $\pm$ 390 & 13000 $\pm$ 290 & 9850 $\pm$ 200 & 12660 $\pm$ 310 \\ 
 & {\fsiii} & 11300 $\pm$ 630 & 15700 $\pm$ 950 & 14300 $\pm$ 900 & 11450 $\pm$ 680 & 14900 $\pm$ 1100 \\ 
 & {\fariii} & 11600 $\pm$ 1450 & $-$ & $-$ & $-$ & $-$ \\ 
 & {\bf $T$(high)} & {\bf 11280 $\pm$ 630} & {\bf 15360 $\pm$ 390} & {\bf 13140 $\pm$ 510} & {\bf 10010 $\pm$ 530} & {\bf 12890 $\pm$ 750} \\
     \hline
     & & POX~4 & SDSS~J1253$-$0312 & Tol~1457$-$262 & He~2$-$10 & \\
     \hline
 {\elecd} ({\cmc})  & {\fsii} & $<$100 & $-$ & 80 $\pm$ 80 &  530 $\pm$ 150 & \\ 
  & {\foii} & 120 $\pm$ 70 & 220 $\pm$ 100 & 150 $\pm$ 70 & 700 $\pm$ 140 & \\
 & {\ffeiii} & 70 - 240 & 200 - 450 & 100 - 1100 & 1400: & \\
 & {\fcliii} &  $-$ &  $-$ & 870$^{+3250}_{-870}$ &  2100$^{+4200}_{-2100}$ & \\
 & {\fariv} & 1690$\pm$950 & $-$ & $-$ & $-$ & \\
 {\elect} (K) &  {\foii} & 11800 $\pm$ 420 & 14920 $\pm$ 930 & 11370 $\pm$ 400 & 8240 $\pm$ 220 & \\
 & {\fsii} & $-$ & $-$ & 11020 $\pm$ 780 & 7050 $\pm$ 310 & \\
 & {\bf $T$(low)} & {\bf 11800 $\pm$ 420} & {\bf 14920 $\pm$ 930} & {\bf 11300 $\pm$ 380} & {\bf 7940 $\pm$ 550} & \\
 & {\foiii} & 12580 $\pm$ 270 & 12840 $\pm$ 330 & 10900 $\pm$ 220 & 7700 $\pm$ 300 & \\
 & {\fsiii} &  18500 $\pm$ 1400 & 14420 $\pm$ 1100 & 12180 $\pm$ 820 & 8390 $\pm$ 330 & \\
 & {\bf $T$(high)} &  {\bf 12580 $\pm$ 270} & {\bf 13000 $\pm$ 570} & {\bf 11000 $\pm$ 420} & {\bf 8040 $\pm$ 420} & \\
     \hline
    \end{tabular} 
   \end{minipage}
  \end{table*}

Physical conditions were calculated from emission-line ratios of CELs with {\sc pyneb} \citep{Luridianaetal12}, an updated {\sc python} version of the {\sc nebular} package of {\sc iraf}, in combination with the atomic data listed in Table~\ref{atomic}. Electron temperature and density, {\elect} and {\elecd}, were determined making use of the diagnostic ratios available in our spectrum that were numerous due to their long exposure times and wide spectral coverage. 
Electron densities have been derived from {\fni}~5198/5200, {\foii}~3726/3729, {\fsii}~6717/6731, {\fcliii}~5518/5538, {\fariv}~4711/4740, and the ratios of several {\ffeiii} lines. Electron temperatures have been calculated using several sets of auroral to nebular line intensity ratios: {\fnii}~5755/(6548+6584), {\foii}~(7319+7330)/(3726+3729), {\fsii}~(4069+4076)/(6717+6731), 
{\foiii}~4363/(4959+5007), {\fsiii}~6312/(9069+9532), and {\fariii}~5192/(7136+7751). {\elect}({\foii}) and {\elect}({\fnii}) have been corrected from the contribution to the intensity of {\foii} 7319, 7330~{\AA} and {\fnii} 5755~{\AA} lines due to recombination following the formulae derived by \citet{liuetal00}. The physical conditions are presented in Table~\ref{physcond}. 
All the objects show low values of {\elecd}, in the range between 100 and 200 cm$^{-3}$ except in the case of He~2$-$10, where the densities 
are between 500 and 2000 cm$^{-3}$. In some objects, the {\elecd} derived from 
{\ffeiii}, {\fcliii}, or {\fariv} show values larger than the other diagnostics, indicating possible density structure. Following \citet{ferlandetal12}, the intensity of {\fni} lines are affected by pumping by far-ultraviolet stellar radiation and therefore the high 
value of {\elecd} determined from those lines in the case of HV (NGC~6822) may be incorrect.  
In general, our {\elecd}  values agree with those derived by \cite{gusevaetal11} and other authors for the same objects. In the case of {\elect} we want to remark the high consistency of the values we obtain from the different indicators for the same objects. The values of {\elect} indicated in Table~\ref{physcond} are also in general consistent with those obtained by other authors but there are some discrepancies. The {\elect}({\fsii}) derived by \cite{gusevaetal11} for He~2$-$10, Mrk~1271, NGC~3125, NGC~5408, HV (NGC~6822), and Tol~1924$-$416 are several thousands K lower than the values we obtain, being our determinations much more consistent with {\elect}({\foii}). This may be due to the use of different sets of atomic data for S$^+$. There are also large differences in the 
values of {\elect}({\foii}) that \cite{gusevaetal11} and us obtain for NGC~5408 and SDSS~J1253$-$0312, being our values much larger but also more consistent with the temperatures obtained for other ions. In contrast, we obtain a much larger {\elect}({\fsiii}) for POX~4, 
and this value is clearly inconsistent with the temperatures of the rest of ions. We assume a two-zone approximation for the nebula estimating the representative values of the electron temperature for the zones where low and high-ionization potential ions 
are present, $T$(low) and $T$(high). Those values will be used for determining ionic abundances. $T$(low) is calculated as the mean of {\elect}({\fnii}), {\elect}({\foii}) and {\elect}({\fsii}) weighted by their inverse relative errors. The same procedure has been 
used to derive $T$(high) from {\elect}({\foiii}), {\elect}({\fsiii}) and {\elect}({\fariii}), except in the case of POX~4, where its abnormal high {\elect}({\fsiii}) has not been considered. These temperatures are included in Table~\ref{physcond}. It must be advised that very recent calculations by \cite{storeyetal14} find rather different values of the collision strengths for CELs of \ion{O}{2+} with respect to those calculated by \cite{palayetal12} -- the dataset that we have used in our calculations for that ion. Using 
the collision strengths of \cite{storeyetal14}  -- the previous collision strengths calculated by \cite{aggarwalkeenan99} give very similar {\elect}({\foiii}) values -- would imply values of {\elect}({\foiii}) of the order of 400-500 K higher than the ones we give in Table~\ref{physcond}.

  \begin{landscape} 
  \begin{table}
   \centering
   \begin{minipage}{230mm}
     \caption{Ionic and total abundances$^{\rm a}$ for HV (NGC~6822), NGC~5408,  Tol~1924$-$416, NGC~3125, and Mkn~1271}
     \label{abundances1}
    \begin{tabular}{lcccccccccc}
     \hline
        & \multicolumn{2}{c}{HV (NGC~6822)} & \multicolumn{2}{c}{NGC~5408} & \multicolumn{2}{c}{Tol~1924$-$416} & \multicolumn{2}{c}{NGC~3125} & \multicolumn{2}{c}{Mkn~1271} \\
     	 & {\tf} = 0.000 & {\tf} = 0.056 & {\tf} = 0.000 & {\tf} = 0.147 & {\tf} = 0.000 & {\tf} = 0.062: & {\tf} = 0.000 & {\tf} = 0.071 & {\tf} = 0.000 & {\tf} = 0.142  \\
	 & &  $\pm$  0.020 & & $\pm$  0.017 & & & & $\pm$  0.020 & & $\pm$  0.031 \\
     \hline
	\multicolumn{11}{c}{Ionic abundances from collisionally excited lines} \\
	\\
	N$^+$ & 5.81 $\pm$ 0.05 & 5.92 $\pm$ 0.06 & 5.48 $\pm$ 0.06 & 5.77 $\pm$ 0.05 & 5.77 $\pm$ 0.05 & 5.88: & 6.26 $\pm$ 0.05 & 6.43 $\pm$ 0.06 & 6.08 $\pm$ 0.07 & 6.58 $\pm$ 0.16 \\
	O$^+$ & 7.22 $\pm$ 0.03 & 7.34 $\pm$ 0.05 & 6.96 $\pm$ 0.04 & 7.26 $\pm$ 0.05 & 7.17 $\pm$ 0.03 & 7.29: & 7.46 $\pm$ 0.06 & 7.64 $\pm$ 0.07 & 7.57 $\pm$ 0.04 & 8.12 $\pm$ 0.20 \\
	O$^{2+}$ & 8.07 $\pm$ 0.06 & 8.31 $\pm$ 0.05 & 7.73 $\pm$ 0.03 & 8.16 $\pm$ 0.04 & 7.92 $\pm$ 0.04 & 8.12: & 8.23 $\pm$ 0.07 & 8.65 $\pm$ 0.08 & 7.96 $\pm$ 0.06 & 8.58 $\pm$ 0.15 \\
	Ne$^{2+}$ & 7.38 $\pm$ 0.08 & 7.64 $\pm$ 0.12 & 7.10 $\pm$ 0.04 & 7.55 $\pm$ 0.07 & 7.26 $\pm$ 0.06 & 7.48: & 7.60 $\pm$ 0.09 & 8.06 $\pm$ 0.17 & 7.35 $\pm$ 0.08 & 8.02 $\pm$ 0.25 \\
	S$^+$ &  5.32 $\pm$ 0.04 & 5.43 $\pm$ 0.05 & 5.23 $\pm$ 0.04 & 5.52 $\pm$ 0.05 & 5.40 $\pm$ 0.04 & 5.51: & 5.64 $\pm$ 0.04 & 5.81 $\pm$ 0.06 & 5.74 $\pm$ 0.06 & 6.24 $\pm$ 0.16 \\
	S$^{2+}$ & 6.43 $\pm$ 0.09 & 6.69 $\pm$ 0.12 & 5.94 $\pm$ 0.04 & 6.40 $\pm$ 0.07 & 6.05 $\pm$ 0.06 & 6.27: & 6.46 $\pm$ 0.10 & 6.93 $\pm$ 0.18 & 6.10 $\pm$ 0.09 & 6.78 $\pm$ 0.26 \\
	Cl$^{2+}$ & 4.51 $\pm$ 0.07 & 4.74 $\pm$ 0.11 & 4.02 $\pm$ 0.05 & 4.44 $\pm$ 0.07 & 4.25 $\pm$ 0.09 & 4.44: & 4.58 $\pm$ 0.11 & 4.98 $\pm$ 0.15 & 4.21 $\pm$ 0.09 & 4.81 $\pm$ 0.21 \\
	Cl$^{3+}$ & 3.89 $\pm$ 0.25 & 4.19 $\pm$ 0.19 & 3.88 $\pm$ 0.09 & 4.25 $\pm$ 0.07 & 3.76 $\pm$ 0.11 & 3.93: & $-$ & $-$ & $-$ & $-$ \\
	Ar$^{2+}$ & 5.90 $\pm$ 0.06 & 6.10 $\pm$ 0.09 & 5.45 $\pm$ 0.03 & 5.82 $\pm$ 0.07 & 5.53 $\pm$ 0.04 & 5.70: & 5.82 $\pm$ 0.11 & 6.18 $\pm$ 0.14 & 5.60 $\pm$ 0.06 & 6.14 $\pm$ 0.19 \\
	Ar$^{3+}$ & 4.75 $\pm$ 0.08 & 4.99 $\pm$ 0.11 & 5.04 $\pm$ 0.03 & 5.48 $\pm$ 0.07 & 5.02 $\pm$ 0.06 & 5.22: & 4.96 $\pm$ 0.10 & 5.39 $\pm$ 0.16 & 5.07 $\pm$ 0.08 & 5.70 $\pm$ 0.22 \\
	Fe$^{2+}$ & 4.91 $\pm$ 0.19 & 5.16 $\pm$ 0.15 & 4.76 $\pm$ 0.08 & 5.20 $\pm$ 0.08 & 5.33 $\pm$ 0.09 & 5.54: & 5.59 $\pm$ 0.24 & 6.02 $\pm$ 0.18 & 5.08 $\pm$ 0.12 & 5.71 $\pm$ 0.23 \\
	\\
	\multicolumn{11}{c}{Ionic abundances from recombination lines} \\
	\\
	He$^+$ & \multicolumn{2}{c}{10.90 $\pm$ 0.01} & \multicolumn{2}{c}{10.90 $\pm$ 0.01} & \multicolumn{2}{c}{10.86 $\pm$ 0.01} & \multicolumn{2}{c}{10.89 $\pm$ 0.02} & \multicolumn{2}{c}{10.85 $\pm$ 0.02} \\
	He$^{2+}$ & \multicolumn{2}{c}{$-$} & \multicolumn{2}{c}{8.98 $\pm$ 0.02} & \multicolumn{2}{c}{8.87 $\pm$ 0.03} & \multicolumn{2}{c}{$-$} & \multicolumn{2}{c}{8.46 $\pm$ 0.08}  \\
	C$^{2+}$ & \multicolumn{2}{c}{7.66 $\pm$ 0.09} & \multicolumn{2}{c}{7.32 $\pm$ 0.14} & \multicolumn{2}{c}{$-$} & \multicolumn{2}{c}{8.03:} & \multicolumn{2}{c}{7.84:} \\
	O$^{2+}$ & \multicolumn{2}{c}{8.31 $\pm$ 0.04} & \multicolumn{2}{c}{8.16 $\pm$ 0.04} & \multicolumn{2}{c}{8.12:} & \multicolumn{2}{c}{8.65 $\pm$ 0.08} & \multicolumn{2}{c}{8.58 $\pm$ 0.15} \\
	\\
	\multicolumn{11}{c}{Total abundances}\\
     	\\
	He & \multicolumn{2}{c}{10.91 $\pm$ 0.01} & \multicolumn{2}{c}{10.91 $\pm$ 0.01} & \multicolumn{2}{c}{10.86 $\pm$ 0.01} & \multicolumn{2}{c}{10.90 $\pm$ 0.02} & \multicolumn{2}{c}{10.85 $\pm$ 0.02} \\
	C & \multicolumn{2}{c}{7.83 $\pm$ 0.10} & \multicolumn{2}{c}{7.59 $\pm$ 0.15} & \multicolumn{2}{c}{$-$} & \multicolumn{2}{c}{8.18:} & \multicolumn{2}{c}{8.27:} \\
	N & 6.72 $\pm$ 0.06 & 6.94 $\pm$ 0.09 & 6.32 $\pm$ 0.08 & 6.72 $\pm$ 0.08 & 6.59 $\pm$ 0.06 & 6.77: & 7.10 $\pm$ 0.08 & 7.48 $\pm$ 0.12 & 6.62 $\pm$ 0.08 & 7.17 $\pm$ 0.30 \\
	O & 8.13 $\pm$ 0.02 & 8.35 $\pm$ 0.05 & 7.80 $\pm$ 0.02 & 8.21 $\pm$ 0.04 & 7.99 $\pm$ 0.02 & 8.18: & 8.30 $\pm$ 0.02 & 8.69 $\pm$ 0.08 & 8.11 $\pm$ 0.02 & 8.71 $\pm$ 0.13 \\
	Ne & 7.44 $\pm$ 0.10 & 7.68 $\pm$ 0.14 & 7.17 $\pm$ 0.05 & 7.61 $\pm$ 0.09 & 7.33 $\pm$ 0.08 & 7.54: & 7.67 $\pm$ 0.12 & 8.10 $\pm$ 0.21 & 7.50 $\pm$ 0.10 & 8.15 $\pm$ 0.34 \\
	S & 6.62 $\pm$ 0.08 & 6.90 $\pm$ 0.12 & 6.16 $\pm$ 0.03 & 6.59 $\pm$ 0.07 & 6.27 $\pm$ 0.05 & 6.47: & 6.66 $\pm$ 0.09 & 7.18 $\pm$ 0.17 & 6.32 $\pm$ 0.07 & 6.92 $\pm$ 0.22 \\
	Ar & 5.94 $\pm$ 0.06 & 6.14 $\pm$ 0.09 & 5.60 $\pm$ 0.02 & 5.99 $\pm$ 0.05 & 5.66 $\pm$ 0.03 & 5.83: & 5.89 $\pm$ 0.10 & 6.25 $\pm$ 0.12 & 5.75 $\pm$ 0.05 & 6.30 $\pm$ 0.16 \\
	Fe & 5.70 $\pm$ 0.21 & 6.05 $\pm$ 0.17 & 5.49 $\pm$ 0.09 & 6.03 $\pm$ 0.10 & 6.05 $\pm$ 0.10 & 6.31: & 6.32 $\pm$ 0.27 & 6.94 $\pm$ 0.22 & 5.54 $\pm$ 0.14 & 6.22 $\pm$ 0.36 \\
     \hline
    \end{tabular}
    \begin{description}
      \item[$^{\rm a}$] In units of 12+log(\ion{X}{+n}/\ion{H}{+}).  
    \end{description}
   \end{minipage}
  \end{table}
  \end{landscape} 

\subsection{Ionic abundances and abundance discrepancy} \label{ionic}

Ionic abundances of \ion{N}{+}, \ion{O}{+}, \ion{O}{2+}, \ion{Ne}{2+}, \ion{S}{+}, \ion{S}{2+}, \ion{Cl}{2+},  \ion{Cl}{3+},  \ion{Ar}{2+}, \ion{Ar}{3+}, and \ion{Fe}{2+} have been derived from 
CELs under the two-zone scheme and not considering temperature fluctuations in the gas ({\tf} $=$ 0, see below) using {the {\sc pyneb} package (version 0.9.3) and the atomic dataset indicated in Table~\ref{atomic}. 
In {\sc pyneb}, an atom is represented as an $n$-level system and computes the line emissivities as a function of the physical conditions, {\elecd} and {\elect}. The final results are presented in Tables~\ref{abundances1} and \ref{abundances2}. 
  We have assumed $T$(low) to calculate the abundances of low ionization potential 
  ions: \ion{N}{+}, \ion{O}{+}, and \ion{S}{+}; and $T$(high) for the rest. 
  We have adopted a mean representative {\elecd} for all the ions in each object. All of them are in the low-density regime, therefore the results do not depend on the precision of the 
  adopted {\elecd}. A number of {\ffeii} lines have been detected in the spectra of some of our sample objects. These lines are severely affected by continuum fluorescence effects 
  \citep[see][]{rodriguez99,verneretal00} and are not suitable for abundance determinations. Unfortunately, the brightest and less sensitive to fluorescence {\ffeii} line, {\ffeii} 
  8616~{\AA} is in one of our observational gaps. In Figure~\ref{o2CELs} we compare the \ion{O}{2+} abundances we determined from CELs and those calculated 
  by \cite{gusevaetal11} from the same dataset for the same objects. In the figure we have also included values from VLT data for the {\hii} galaxy NGC~5253, which analysis 
  was published by \cite{lopezsanchezetal07}. 
  These authors used a similar methodology as in this paper so their data can be considered as part of this work. Although the areas extracted by \cite{gusevaetal11} and us for 
  their spectroscopical analysis may be different in some objects -- perhaps in the case of NGC~5408 -- there is a general good agreement because most of the points lie very close of the 1:1 relation line. 
  This is expected for independent similar calculations based on the same data. It is remarkable that since the error bars of all our abundance determinations --except NGC~5408-- intersect the 1:1 relation line, none of the error bars 
  quoted by \cite{gusevaetal11} intersect it at all. This fact suggests that the uncertainties calculated by those authors might be underestimated.

  \begin{table*}
   \centering
   \begin{minipage}{180mm}
     \caption{Ionic and total abundances$^{\rm a}$ for POX~4, SDSS~J1253$-$0312, Tol~1457$-$262, and He~2$-$10}
     \label{abundances2}
    \begin{tabular}{lccccc}
     \hline
        & \multicolumn{2}{c}{POX~4} & SDSS~J1253$-$0312 & Tol~1457$-$262 & He~2$-$10 \\
     	 & {\tf} = 0.000 & {\tf} = 0.060: & {\tf} = 0.000 & {\tf} = 0.000 & {\tf} = 0.000  \\
     \hline
	\multicolumn{6}{c}{Ionic abundances from collisionally excited lines} \\
	\\
	N$^+$ & 5.77 $\pm$ 0.06 & 5.91: & 5.96 $\pm$ 0.07 & 6.15 $\pm$ 0.05 & 7.53 $\pm$ 0.09 \\
	O$^+$ & 7.23 $\pm$ 0.03 & 7.38: & 6.81 $\pm$ 0.04 & 7.58 $\pm$ 0.03 & 8.41 $\pm$ 0.10 \\
	O$^{2+}$ & 8.03 $\pm$ 0.03 & 8.24: & 7.96 $\pm$ 0.05 & 8.10 $\pm$ 0.05 & 8.00 $\pm$ 0.08 \\
	Ne$^{2+}$ & 7.35 $\pm$ 0.04 & 7.58: & 7.32 $\pm$ 0.06 & 7.46 $\pm$ 0.08 & 7.16 $\pm$ 0.14 \\
	S$^+$ &  5.45 $\pm$ 0.07 & 5.59: & 5.29 $\pm$ 0.09 & 5.84 $\pm$ 0.04 & 6.30 $\pm$ 0.10 \\
	S$^{2+}$ & 6.17 $\pm$ 0.14 & 6.40: & 6.10 $\pm$ 0.07 & 6.35 $\pm$ 0.07 & 6.77 $\pm$ 0.08 \\
	Cl$^{2+}$ & 4.21 $\pm$ 0.09 & 4.41: & 4.08 $\pm$ 0.12 & 4.65 $\pm$ 0.14 & $-$ \\
	Ar$^{2+}$ & 5.54 $\pm$ 0.05 & 5.72: & 5.57 $\pm$ 0.06 & 5.79 $\pm$ 0.05 & 6.01 $\pm$ 0.07 \\
	Ar$^{3+}$ & 5.17 $\pm$ 0.06 & 5.38: & 5.21 $\pm$ 0.09 & 5.14 $\pm$ 0.11 & $-$ \\
	Fe$^{2+}$ & 5.23 $\pm$ 0.12 & 5.45: & 5.24 $\pm$ 0.09 & 5.73 $\pm$ 0.10 & 6.21 $\pm$ 0.10 \\
	\\
	\multicolumn{6}{c}{Ionic abundances from recombination lines} \\
	\\
	He$^+$ & \multicolumn{2}{c}{10.85 $\pm$ 0.02} & 10.90 $\pm$ 0.02 & 10.89 $\pm$ 0.02 & 10.74 $\pm$ 0.02  \\
	He$^{2+}$ & \multicolumn{2}{c}{9.03 $\pm$ 0.03} & 8.82 $\pm$ 0.04 & 8.70 $\pm$ 0.07 & $-$  \\
	C$^{2+}$ & \multicolumn{2}{c}{8.02:} & 7.61: & $-$ & $-$ \\
	O$^{2+}$ & \multicolumn{2}{c}{8.24:} & 7.86: & $-$ & $-$ \\
	\\
	\multicolumn{6}{c}{Total abundances}\\
     	\\
     	He & \multicolumn{2}{c}{10.86 $\pm$ 0.02} & 10.90 $\pm$ 0.02 & 10.89 $\pm$ 0.02 & $-$  \\
	C & \multicolumn{2}{c}{8.11:} & 7.70: & $-$ & $-$ \\
	N & 6.63 $\pm$ 0.07 & 6.83: & 7.14 $\pm$ 0.08 & 6.78 $\pm$ 0.06 & 7.67 $\pm$ 0.14 \\
	O & 8.09 $\pm$ 0.02 & 8.30: & 7.99 $\pm$ 0.02 & 8.21 $\pm$ 0.02 & 8.55 $\pm$ 0.02 \\
	Ne & 7.41 $\pm$ 0.05 & 7.63: & 7.35 $\pm$ 0.08 & 7.57 $\pm$ 0.10 & 7.71 $\pm$ 0.16 \\
	S & 6.39 $\pm$ 0.12 & 6.67: & 6.41 $\pm$ 0.06 & 6.55 $\pm$ 0.05 & 6.90 $\pm$ 0.08 \\
	Ar & 5.70 $\pm$ 0.04 & 5.89: & 5.73 $\pm$ 0.05 & 5.90 $\pm$ 0.05 & 6.31 $\pm$ 0.22 \\
	Fe & 5.98 $\pm$ 0.13 & 6.25: & 6.28 $\pm$ 0.11 & 6.28 $\pm$ 0.12 & 6.34 $\pm$ 0.16 \\
     \hline
    \end{tabular}
    \begin{description}
      \item[$^a$] In units of 12+log(\ion{X}{+n}/\ion{H}{+}).  
    \end{description}
   \end{minipage}
  \end{table*}

  \begin{figure}
   \centering
   \includegraphics[scale=0.42]{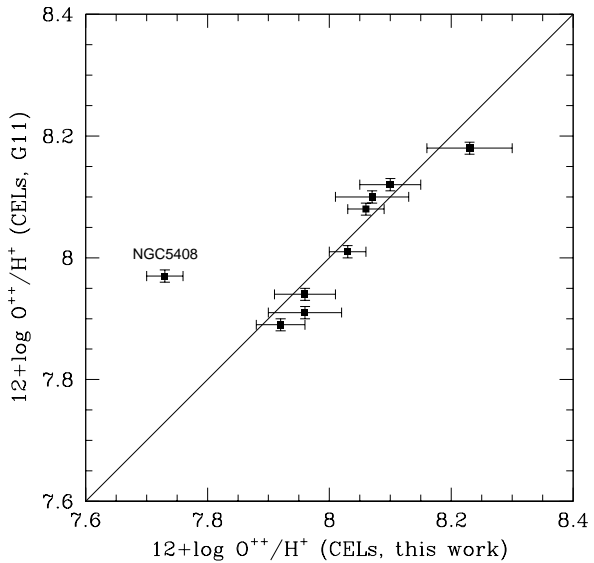} 
   \caption{Comparison of the  \ion{O}{2+} abundances we determine from collisionally excited lines (CELs) and those determined by \citet{gusevaetal11} from the same dataset for the 
   same objects. The continuous line is the 1:1 relation. Note that our quoted errors are consistent --as expected-- with the 1:1 relation. In contrast, the errors quoted 
   by \citet{gusevaetal11} are much smaller and not consistent with that relation. The large discrepancy for NGC~5408 is due to the different areas of the object extracted in both 
   studies. We include values from VLT spectra of the {\hii} galaxy NGC~5253 published by \citet{lopezsanchezetal07} as part of this work (average values of their zones A and B).}
   \label{o2CELs}
  \end{figure}    	 

We have measured many {\hei} emission lines in the spectra of all objects. To derive the He$^+$ abundance, we used the effective recombination coefficients of \cite{storeyhummer95} for {\hi} and those computed by 
\cite{porteretal12,porteretal13} for {\hei}, whose calculations include corrections for collisional excitation and self-absorption effects. The finally adopted He$^+$/H$^+$ ratio for each 
object has been derived from a maximum likelihood method \citep[MLM, ][]{peimbertetal00, apeimbertetal02}. With this method we obtain the best pair of values of mean He$^+$/H$^+$ 
ratio and temperature fluctuations parameter -- {\ts}, in this case we designate it as {\ts(\hei)} -- that minimize the $\chi^2$ of the MLM. For that procedure, we used a number of {\hei} 
lines ranging from 5 to 15 depending on the object. Table~\ref{t2} shows the values of {\ts(\hei)} we obtain for 
all the objects and those estimated assuming that the abundance discrepancy factor for \ion{O}{2+} -- {\adfo}, see definition in Equation 1 -- is produced by temperature fluctuations in the ionised gas (see below). As we can see the {\ts} values we 
obtain from both methods are in general consistent except in the case of NGC~5408. It is clear that since the {\ts} parameter determined from the {\adfo} concerns to the zone of the nebula 
where \ion{O}{2+} is present and {\ts(\hei)} is representative for the whole ionized gas. However, in the sample objects, the \ion{O}{2+}/\ion{O}{+} ratio is between 2.5 and 7.0, indicating that both zones have a large common volume. 
The only object with previous determination of {\ts} is HV in NGC~6822.  
\cite{apeimbertpeimbert05} obtain a value of 0.076$\pm$0.018 for it based on the {\adfo}, consistent with our {\ts} determinations within the uncertainties. The values of the {\ts} we calculate in the objects investigated in this paper are among the 
highest we have found in {\hii} regions. In the compilations presented by \cite{lopezsanchezetal07} and \cite{estebanetal09}, we can see that only NGC~2363 -- a giant {\hii} region in the dwarf {\hii} galaxy NGC~2366 -- for which \cite{estebanetal09} 
determined a {\ts} of about 0.120, show a similar high value. Therefore, giant {\hii} regions in dwarf star-forming galaxies present the highest {\ts} values. This may perhaps be related to their complex structure and dynamics 
due to the action of stellar winds and supernova remnants as it was firstly advocated by \cite{peimbertetal91} or even the hardening of the energy distribution of ionizing photons as the central clusters evolve \citep{perez97}. Although the simple Peimbert's formalism \citep{peimbert67} may be not completely adequate for such large temperature fluctuations, we consider it is still useful for parameterizing the problem until their existence  
and nature are definitely proven.

  We have obtained good signal-to-noise ratio measurements of the intensity of {\cii} 4267~{\AA} line in HV (NGC~6822) and NGC~5408. In the cases of Mkn~1271, NGC~3125, POX4, and SDSSS~J1253$-$0312 the error of the intensity of that line is of the order or larger than 40\%, and does not permit to derive a precise value of the ionic abundance. {\cii} 4267~{\AA}  has not been detected in Tol~1457$-$262 and He~2$-$10. However, \cite{gusevaetal11} report good measurements of the intensity of {\cii} 4267~{\AA} in all the sample objects except in
SDSSS~J1253$-$0312  -- its error is somewhat larger than 40\%  --  and He~2$-$10, for which they derive an upper limit. This disagreement has not a clear explanation for us considering that we are analysing the same spectral dataset and that 
-- as it has been said before -- we optimized the extracted areas in order to isolate the brightest knots for each object. These knots coincide with the position of the ionizing sources and therefore where the bulk 
of \ion{C}{2+} emission should be concentrated.

The lower panel of Figure~\ref{c2rls} compares our line intensity ratios of {\cii} 4267~{\AA} with respect to H$\beta$  with those obtained by \cite{gusevaetal11}. As we can see, some values do not agree, but there is not a systematical difference between both sets of determinations. {\cii} 4267~{\AA}  is a pure RLs and permits to derive the \ion{C}{2+}/\ion{H}{+} ratio. We have used $T$(high) and the effective recombination coefficients of \cite{daveyetal00} to obtain the \ion{C}{2+} abundance of the objects, which is included in Tables~\ref{abundances1} and \ref{abundances2}. The upper panel of Figure~\ref{c2rls} shows that the \ion{C}{2+}/ \ion{H}{+} ratios we obtain for the sample objects  -- including data for NGC~5253 
\citep{lopezsanchezetal07} -- are systematicaly larger than those determined by \cite{gusevaetal11} except in the case of NGC~5408. The fact that this systematic difference is not seen in the lower panel and that we are using the same set of effective recombination coefficients suggests that the source of the disagreement should come from the different -- or inappropriate -- calculation procedure used to derive the abundances. 

  \begin{table}
  \centering
   \begin{minipage}{83mm}
     \caption{Values of {\ts}.}
     \label{t2}
    \begin{tabular}{lcc}
     \hline
     Object &  \ts(ADF)$^{\rm a}$ & \ts(\hei)$^{\rm b}$  \\
     \hline
     He 2$-$10 & $-$ & 0.020$^{+0.054}_{-0.020}$ \\
     Mkn~1271 & 0.142$\pm$0.031 & 0.167$\pm$0.026 \\
     NGC~3125 & 0.071$\pm$0.020 & 0.081$\pm$0.036 \\
     NGC~5408 & 0.147$\pm$0.017 & 0.063$\pm$0.034 \\
     HV (NGC~6822) & 0.056$\pm$0.020 & 0.080$\pm$0.030 \\
     POX~4 & 0.060:& 0.068$\pm$0.048 \\
     SDSS~J1253$-$0312 &  $-$ & 0.052$^{+0.064}_{-0.052}$ \\
     Tol~1457$-$262 & $-$ & 0.049$^{+0.052}_{-0.049}$ \\
     Tol~1924$-$416 & 0.062:& 0.116$\pm$0.033 \\
     \hline
    \end{tabular}
    \begin{description}
      \item[$^{\rm a}$] {\ts} estimated from the {\adfo}.  
      \item[$^{\rm b}$] {\ts} estimated from the MLM used for determining the He$^+$/H$^+$ ratio.       
    \end{description}
   \end{minipage}
  \end{table}

  \begin{figure}
   \centering
   \includegraphics[scale=0.85]{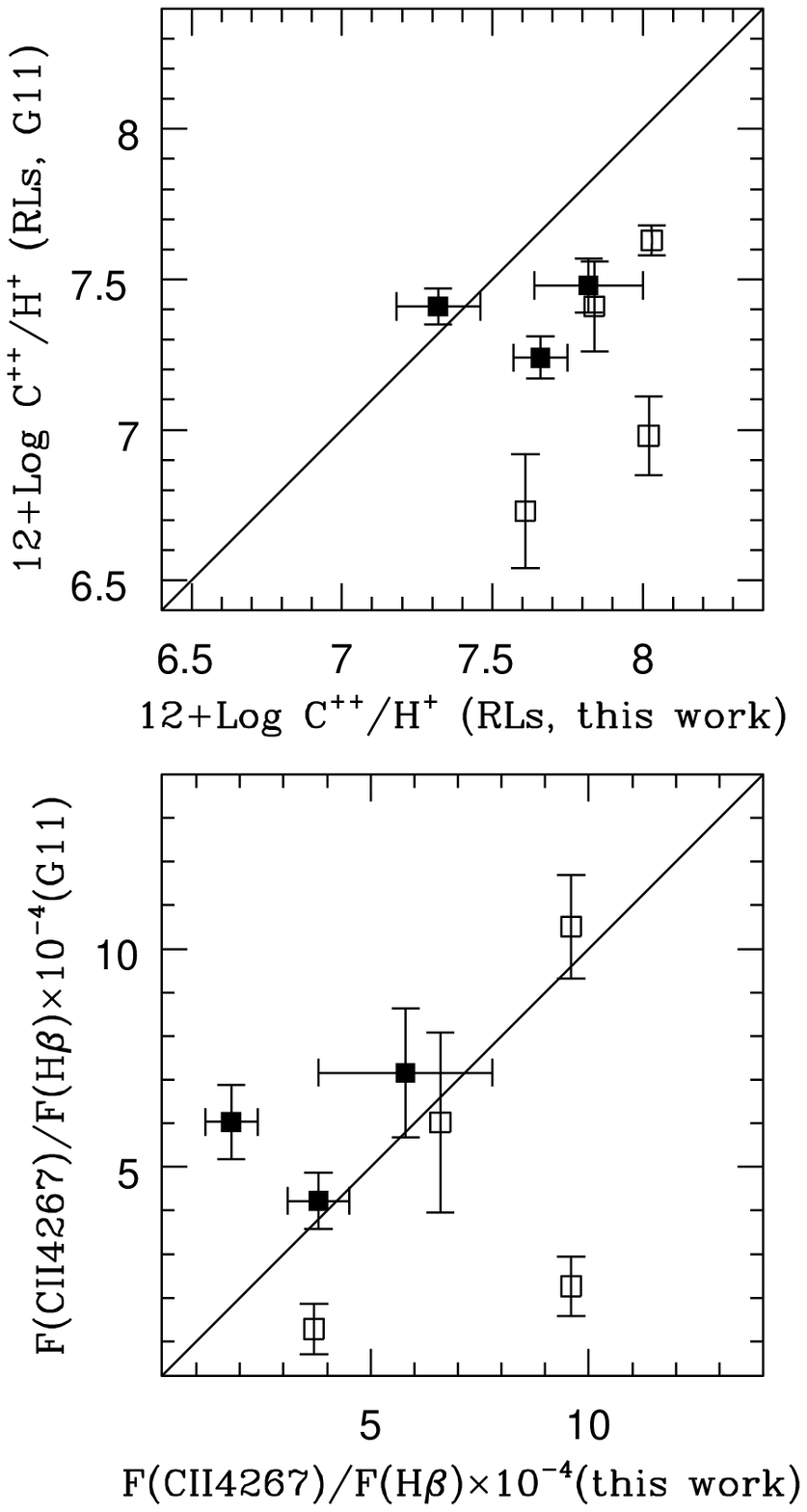} 
   \caption{Upper panel: comparison of the \ion{C}{2+} abundances we determine from the {\cii} 4267~{\AA} recombination line (RL) and those calculated by \citet{gusevaetal11} from the   same dataset for the same objects. The solid lines represents the 1:1 relation. The filled squares represent objects for which we have good determinations of the intensity of the {\cii} 4267~{\AA} line, open squares are objects for which we report  errors larger than 40\% in the intensity of that line. We include values from VLT spectra of the {\hii} galaxy NGC~5253 published by \citet{lopezsanchezetal07} as part of the sample of this work. Lower panel: the same but comparing derredened line intensity ratios of {\cii} 4267~{\AA}  with respect to H$\beta$. }
   \label{c2rls}
  \end{figure}    	 
 
  We obtain good measurements of RLs of {\oii} in the cases of HV (NGC~6822),  Mkn~1271, NGC~3125, and NGC~5408 and with errors of the order or larger than 40\% in POX~4, Tol~1924$-$416, and SDSS~J1253$-$0312. \cite{gusevaetal11} obtain good 
  measurements -- errors in the line intensities always better than 35\% -- in He~2$-$10, HV (NGC~6822),  Mkn~1271, NGC~3125, NGC~5408, and POX~4. It is striking that \cite{gusevaetal11} give an error of 15\% for the blend of 
 {\oii} lines at 4649.13 and 4650.84~{\AA} in He~2$-$10 since we do not detect such feature in our spectrum. In Figure~\ref{o2rls} we compare the line intensity ratios of the blend of 
 {\oii} 4649.13 and 4650.84~{\AA} lines with respect to H$\beta$ obtained by \cite{gusevaetal11} and in this work, finding no systematical differences. Following our usual methodology to minimize uncertainties, we have derived the \ion{O}{2+} abundance from the estimated total intensity of 
 the RLs of multiplet 1 of {\oii} \citep[see Equation 7 of][]{estebanetal09}. The lines of multiplet 1 of {\oii} are not in LTE for densities {\elecd} $<$ 10$^4$~cm$^{\rm -3}$ \citep{ruizetal03}, and this is the case for all our sample objects. Therefore, we have used the prescriptions given by \citet{apeimbertetal05} to calculate the appropriate corrections for the relative strengths between the individual lines of multiplet 1. 
The \ion{O}{2+} abundances from RLs have been calculated using $T$(high) and the effective recombination coefficients of \cite{storey94} assuming LS coupling and they are 
included in Tables~\ref{abundances1} and \ref{abundances2}. 
In contrast to that we found in the comparison of  \ion{C}{2+} abundances, we do not find systematic differences between the \ion{O}{2+}/\ion{H}{+} ratios determined by us and the values given by \cite{gusevaetal11}. In this case, there seems to be no discrepancies between the methods for determining the \ion{O}{2+} abundance used by both research groups.
 
The spectra of some of the sample objects also show permitted lines of other heavy-element ions as {\nitroi}, {\nii}, {\niii}, {\oi} and/or {\silii}, but they are produced by fluorescence and can not be used to derive reliable abundance values \citep[see][]{estebanetal98,estebanetal04}.  

  \begin{figure}
   \centering
   \includegraphics[scale=0.85]{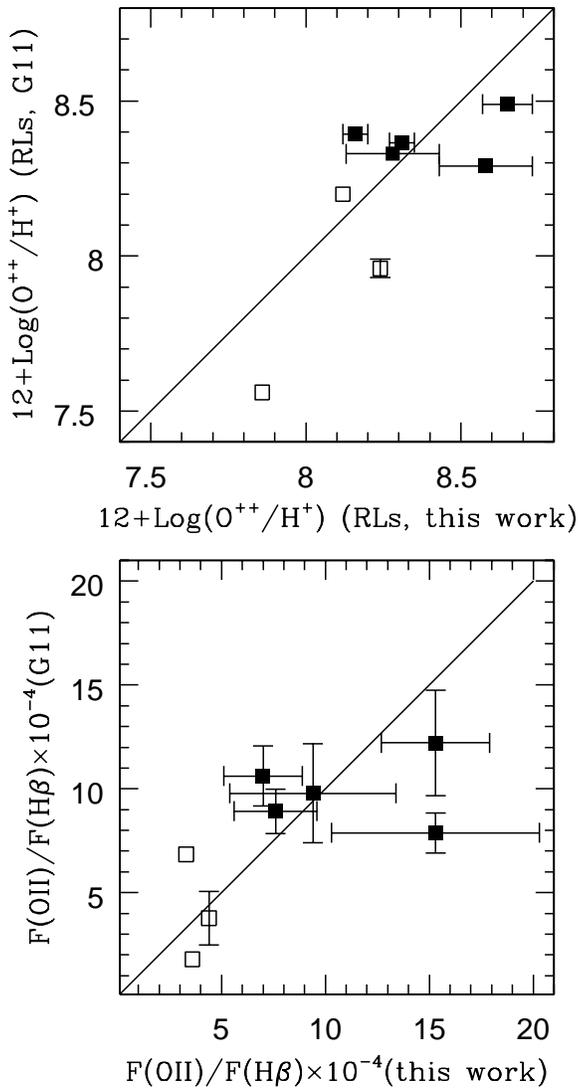} 
   \caption{Comparison of the dereddened  intensity ratios of the {\oii} 4649.13 + 4650.84~{\AA} lines with respect to H$\beta$ and those obtained by \citet{gusevaetal11} from the same dataset for the same objects. The solid line represents the 1:1 relation. The filled squares represent objects for which we have good determinations of the intensity of the {\oii} lines, open squares are objects for which we report errors larger than 40\%. We include values from VLT spectra of the {\hii} galaxy NGC~5253 published by \citet{lopezsanchezetal07} as part of the sample of this work.}
   \label{o2rls}
  \end{figure}    	 
 
For HV (NGC~6822), Mkn~1271, NGC~3125, and NGC~5408, the high signal-to-noise ratio of the spectra has permitted to calculate \ion{O}{2+} abundances from both kinds of lines RLs and CELs. In the cases of POX~4, Tol~1924$-$416, and SDSS~J1253$-$0312 the abundance determinations based on RLs are rather uncertain due to the faintness of the lines. It is important to remark that -- except for SDSS~J1253$-$0312 -- the \ion{O}{2+} abundance determined from RLs is always larger than that obtained from CELs, with differences ranging from 0.20 and 0.62 dex. This difference is the so-called abundance discrepancy \ion{O}{2+}, {\adfo}, which is defined as: 
\begin{eqnarray}
 {\rm ADF}({\rm O}^{2+}) = {\rm log}({\rm O}^{2+}/{\rm H}^+)_{\rm RLs} - {\rm log}({\rm O}^{2+}/{\rm H}^+)_{\rm CELs}. 
\end{eqnarray} 
Positive values of the {\adfo} have been found in all {\hii} regions where {\oii} RLs lines have been measured with enough signal-to-noise 
ratio \citep[e.g.][]{garciarojasesteban07,estebanetal09}. The mean value of the {\adfo} found for Galactic and extragalactic {\hii} regions 
is 0.26$\pm$0.09 \citep{estebanetal09}. The mean of the four objects with the best determinations of the {\adfo} in the present work (HV, Mkn~1271, NGC~3125 and NGC~5408) is 0.43$\pm$0.16, higher than in the rest of the {\hii} regions reported in the literature.  There seems to be no correlation between the {\adfo} and properties as metallicity, ionization 
degree or mean electron temperature of the ionized gas. 

  If -- as an hypothesis -- and as in other previous spectroscopical studies of our group, we assume the validity of the temperature fluctuations paradigm and that this phenomenon produces 
  the abundance discrepancy \citep[see][]{garciarojasesteban07}, we can estimate the \tf\ parameter 
  that produces the agreement between the abundance of \ion{O}{2+} determined from CELs and RLs, \tf(ADF).  Table~\ref{t2} compiles the values of \tf(ADF) obtained for the objects. 
  Tables~\ref{abundances1} and \ref{abundances2} list the ionic abundances determined for \tf = 0 -- the standard procedure when not considering temperature fluctuations -- and 
  assuming the estimated \tf(ADF) values and their associated uncertainties. These last calculations have been made following the formalism 
  outlined by \cite{peimbertcostero69} and updated by \cite{apeimbertetal02}. 
  
  \subsection{Gas-phase total abundances} \label{total}
  
  We have adopted a set of ionization correction factors (ICFs) to correct for the unseen 
  ionization stages and derive the total gas-phase abundances of the different elements, except in the case of O for 
  which we have measured emission lines of all the main expected ionic species. 
  The adopted total abundances for \tf\ = 0 and  \tf\ $>$ 0 are presented in Tables~\ref{abundances1} and \ref{abundances2}. 
  For those objects without detection of {\heii} lines -- HV (NGC~6822), NGC~3125, and He~2$-$10 -- the total helium abundance has been corrected for the presence of neutral helium using the 
  ICFs obtained from the photoionization models of \cite{stasinska90} that reproduce the O abundance, \ion{O}{2+}/\ion{O}{+} ratio, {\elect}({\foiii}) and {\elect}({\foii}) determined for each object. In the case of 
  He~2$-$10, we have not been able to obtain a precise value of the total He abundance due to its low ionization degree that implies a large contribution of neutral He and a very uncertain ICF. The total He 
  abundance of the rest of the sample objects have been calculated adding the  \ion{He}{+}/\ion{H}{+} and  \ion{He}{2+}/\ion{H}{+} ratios, the contribution of this last ion is almost negligible in all the objects. The 
  weakness of the {\heii} 4686~{\AA} lines implies that the amount of \ion{O}{3+} inside the nebulae is irrelevant, validating our assumption of no considering an ICF for O. 
  
  For C, we have adopted an ICF derived from photoionization models of \cite{garnettetal99}. This correction seems to be fairly appropriate considering the high ionization degree of the nebulae where the {\cii} 4267~{\AA} line 
  has been detected. In order to derive the total abundance of nitrogen we have used the usual ICF based on the similarity of the ionization potential 
  of \ion{N}{+} and \ion{O}{+} \citep{peimbertcostero69}. The only measurable CELs of Ne in the optical range are those of \ion{Ne}{2+} 
  but the fraction of \ion{Ne}{+} may be important in the nebulae. We have adopted the usual expression 
   of \citet{peimbertcostero69} that assumes that the ionization structure of Ne is similar to that of O. We have measured CELs of two ionization stages of 
  S: \ion{S}{+} and \ion{S}{2+}, and used the ICF proposed by \cite{stasinska78} to take into account the 
  presence of some \ion{S}{3+}. 
  For argon, we have determinations of \ion{Ar}{2+} and \ion{Ar}{3+} but some contribution of 
  \ion{Ar}{+} is also expected. We have adopted the ICF recommended by \cite{izotovetal94} 
  for this element. Finally, we have used an ICF scheme based on photoionization models 
  of \cite{rodriguezrubin05} to obtain the total Fe/H ratio from the \ion{Fe}{2+}/ \ion{H}{+} ratio. 
 The variations due to the dependence of the adopted ICFs on the  \tf\ considered are also included  
  in the total abundances given in Tables~\ref{abundances1} and \ref{abundances2}. We have made the exercise of estimating the total abundances 
  of several elements in the case of using other ICFs as the scheme proposed by \cite{izotovetal06} based on photoionization model grids. We find that 
 the differences are not very significant except in rather few cases.  Our determinations of the N and Ne abundances are in average about 0.04 dex higher than those 
 calculated making use of the formulae of \cite{izotovetal06}. In the case of S, the average difference is higher, being our S/H ratios 0.09 dex 
 lower than those obtained with the alternative ICF scheme. The differences are much larger in the cases of SDSS~J1253$-$0312 and He~2$-$10 -- precisely our  
 objects with the highest and lowest ionization degrees, respectively -- for which our S abundances are 0.22 and 0.32 dex -- respectively -- lower 
 than those estimated using the formulation of \cite{izotovetal06}. Finally, our Ar abundances are only about 0.02 dex higher except in the case of He~2$-$10, where 
 the difference is about 0.22 dex.
 
\section{Discussion} \label{discussion}
 
 \subsection{The C/O {\it vs.} O/H relation} \label{covsoh}
 
 Determining the C/O ratios in extragalactic {\hii} regions and {\hii} galaxies permits to explore the chemical evolution of C in low-metallicity objects. Figure~\ref{covso} plots 
 the C/O {\it vs.}~O/H ratios for the complete sample of {\hii} regions where C and O abundances have been determined from RLs: HV (NGC~6822) and NGC~5408 from this work, Galactic {\hii} regions \citep{garciarojasesteban07, estebanetal13}, {\hii} regions in external spiral and irregular galaxies \citep{apeimbert03, estebanetal09} and the blue compact dwarf galaxy NGC~5253 \citep{lopezsanchezetal07}. The peculiar position of NGC~2363 is indicated in Figure~\ref{covso} because it will be discussed later. 
 The figure shows the well-known C/O {\it vs.}~O/H correlation found in previous works by, for example, \citet{garnettetal95, garnettetal99} from observations of CELs or \citet{estebanetal02, estebanetal09} from observations of RLs. 
 
  \begin{figure}
   \centering
   \includegraphics[scale=0.42]{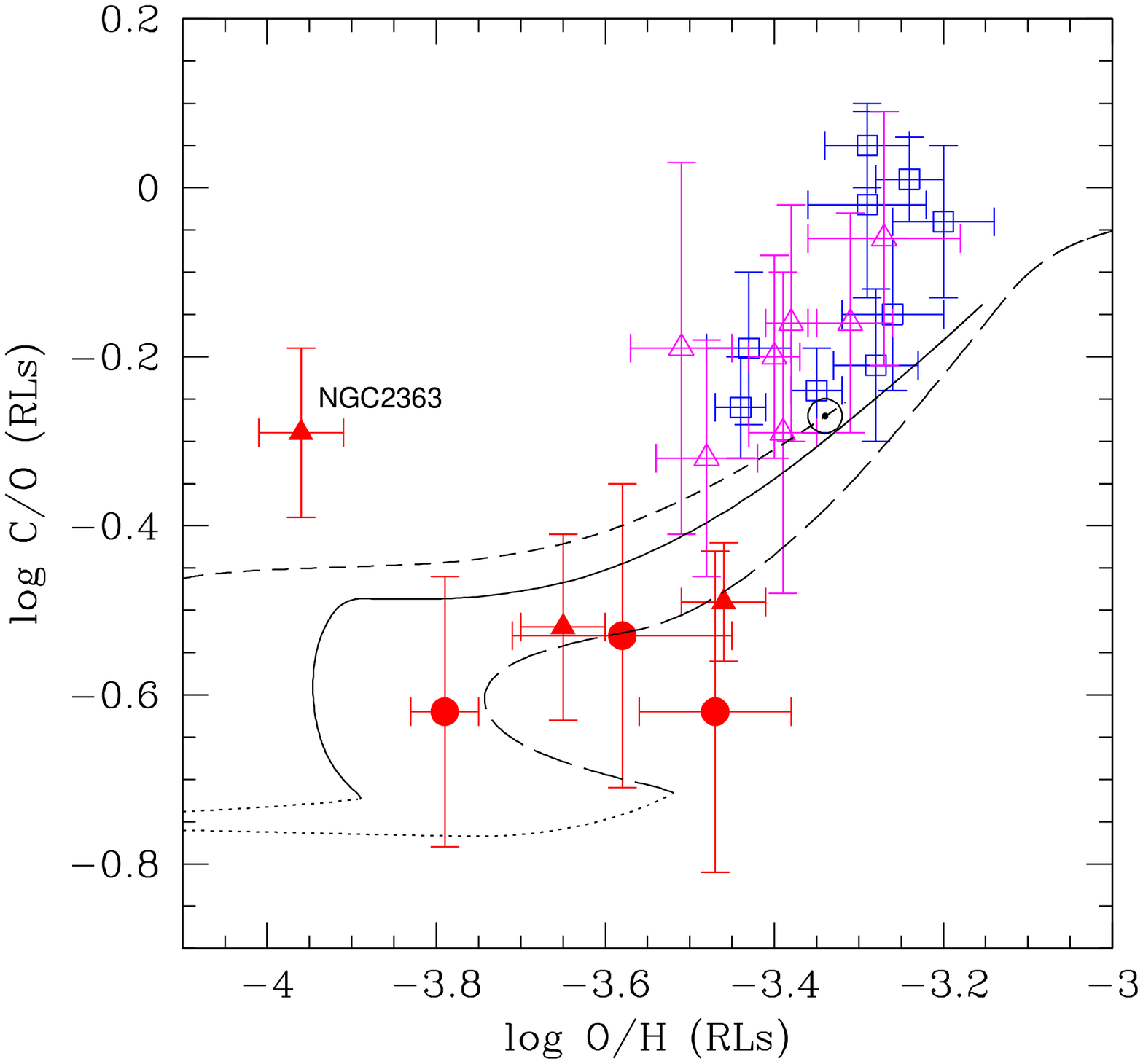} 
   \caption{C/O {\it vs.}~O/H ratios of Galactic and extragalactic \hii\ regions determined from recombination lines. open (blue) squares represent data for Galactic \hii\  regions \citep{garciarojasesteban07, estebanetal13}; open (magenta) triangles correspond to  {\hii} regions in the inner discs of external spiral galaxies \citep{estebanetal09}; full (red) triangles indicate 
   abundance ratios for {\hii} regions in irregular galaxies \citep[this work;][]{apeimbert03,estebanetal09} and full (red) circles correspond to dwarf star-forming galaxies \citep[classified as  {\hii} galaxies or blue compact dwarves: this work;][]{lopezsanchezetal07}. The solar symbol 
  represents the abundances of the Sun \citep{asplundetal09}. The lines show the predictions of chemical evolution models presented by \citet{estebanetal13} 
  for the Milky Way. The solid line represents the time evolution of abundances since the formation of the Galactic disc ($t_G$ $>$ 1 Gy) for $R_G$ = 8 kpc, the long- and short-dashed lines represent the same but for $R_G$ = 4 and 14 kpc, respectively. Dotted lines represent the evolution for the same radii but for $t_G$ $<$ 1 Gy, during the formation of the Galactic halo.  We recommend to add $\sim$0.1 dex to the log(O/H) values of {\hii} regions to correct for dust depletion. No correction is needed for their C/O ratios (see \S\ref{covsoh} for details). }
   \label{covso}
  \end{figure}    	 

Figure~\ref{covso} includes the time evolution of the abundances predicted by the chemical evolution models of the Milky Way disc presented by \citet{estebanetal13}. 
 These models reproduce the observed radial C, O, and C/O gradients determined by \citet{estebanetal05, estebanetal13}. The solid line represents the time evolution 
 of abundances at $R_G$ = 8 kpc since the epoch of the formation of the Galactic disc, from $t_G$ $\sim$ 1 Gy up to the present. The dotted line connecting with the solid line indicates 
 the chemical evolution at that part of the Galaxy for $t_G$ $<$ 1 Gy, corresponding to the epoch of the formation of the Galactic halo. The long-dashed line connected with another 
 dotted line represents the chemical evolution at  $R_G$ = 4 kpc during the halo and disc formation. The short-dashed line represents the same for $R_G$ = 14 kpc. 
 
The main assumptions of the chemical evolution model by \citet{estebanetal13} are described in the following:
 \begin{itemize}
\item[i)] The Milky Way disc was formed in an inside-out scenario from primordial infall with time scales $\tau = r$(kpc)$- 2$ Gyr. 
\item[ii)] The star formation rate (SFR) is a spatial and temporal function of the form:
\begin{equation}
    {\rm SFR}(r,t) = \nu(r) \times M_{\rm bar}^{1.4}(r,t) \times M_{\rm gas}^{0.4}(r, t);
  \end{equation}    
where $M_{\rm gas}$ and $M_{\rm bar}$ are the surface mass density of gas and baryonic masses; $\nu$ is $r$-dependent to reproduce the behaviour of the O/H gradient in the Galactic disk.
\item[iii)] The initial mass function is that by \citet{kroupaetal93} in the $0.08 - 80$ $M_\odot$ range. 
\item[iv)] We consider an array of metal dependent yields:  
	 \begin{itemize}
	 \item[a)] for LIM stars ($0.8 \leq m/M_\odot \leq 8$), we use the yields by \citet{marigoetal96,marigoetal98} and \citet{portinarietal98}; 
	 \item[b)] for massive stars ($8 \leq m/M_\odot \leq 80$) we considered the yields including stellar rotation by \citet{hirschi07} and \citet{meynetmaeder02} for $Z \leq 0.004$; and the intermediate wind yields, obtained as an average of the yields by \citet[][high mass-loss rate]{maeder92} and \citet[][low mass-loss rate]{hirschietal05} for $Z = 0.02$. 
 	\end{itemize}  
 \end{itemize}
 
Our model is in agreement with competing classical chemical evolution models (without dynamical considerations) in broad outlines, but it disagrees on some details. For example, 
our model explains the rise in C/O for metal-poorest thick disk stars due to LIM stars of low $Z$ and the high C/O for metal-richest thin disk stars due to massive stars of high $Z$, 
but \citet{cescuttietal09} explain the C/O ratios in disk stars due to massive stars alone. Most chemical evolution models need to improve their agreement with the gaseous and stellar N abundances \citep{carigietal05,romanoetal10,kobayashietal11}. The disagreement is due mainly to the uncertainties in the N yields \citep{karakaslattanzio07} due to its complex nucleosynthesis. 
Moreover, our model, as classical ones, cannot reproduce the alpha-element enhancement shown by the thick disk stars compared to thin disk ones  
because it does not consider dynamical aspects \citep[see][and references therein]{minchevetal13}.

It must be remarked that the C and O abundances of the \hii\ regions presented in Figure~\ref{covso} have not been corrected for the fraction of atoms embedded
in dust and this should be taken into account when comparing with stellar or solar abundances or the results from chemical evolution models. 
According to \cite{mesadelgadoetal09} the fraction of O 
embedded in dust grains in the Orion Nebula is about 0.12 dex. \cite{apeimbertpeimbert10} have estimated that the depletion of O increases with increasing O/H. They propose that O depletion ranges from about 0.08 dex, for the metal poorest {\hii} regions, to about 0.12 dex -- the value estimated for the Orion Nebula -- for metal-rich {\hii} regions. 
C is also expected to be depleted in dust, mainly in polycyclic aromatic hydrocarbons and graphite, though the study of its depletion is certainly problematic \citep[see][]{mathis96}. 
The paucity of observational information makes that the C abundance in the gas and dust phases of the diffuse interstellar medium is something very poorly understood \citep[see][and references therein]{sofiaetal11}. The different estimations of the amount of C locked up in dust grains in the local interstellar medium show a rather large variation depending on the abundance determination methods applied \citep[see][]{jenkins14}. 
Assuming the quantities complied by \cite{jenkins14} we find that the gas+dust C/H ratios in the Orion Nebula would be between 0.09 to 0.26 dex larger than the ionized gas phase values determined by \cite{estebanetal04}. 
However, some degree of destruction of C-bearing dust particles is likely to be acting in ionized nebulae and the assumption of depletion factors of the neutral diffuse interstellar medium may not be 
appropriate for an ionized nebula. On the other hand, assuming the protosolar C abundance given by \cite{asplundetal09} -- corrected for the amount of this element produced by the chemical evolution of the Galaxy that  those authors propose -- we find a dust depletion of the order of 0.10 dex for the Orion Nebula, in agreement with the estimate by \cite{estebanetal98}. 
Taking into account (a) the uncertain degree of depletion of C in local objects, (b) the indications that this seems to be not very different to that of O and (c) the complete lack of estimations for lower metallicities, we 
consider reasonable to assume no correction for the C/O ratio. As a conclusion, we recommend to increase nebular 
O/H values by about 0.08-0.12 dex but not the C/O ratio when comparing those ratios with the predictions of chemical evolution models or solar or stellar abundances.
 
Following Figure~\ref{covso}, it is evident that  \hii\ regions in irregular and {\hii} galaxies -- except NGC~2363 -- show systematically lower values of C/O and somewhat lower O/H than \hii\ regions belonging to the inner discs of spiral galaxies. Therefore, both kinds of objects occupy different loci in the diagram, indicating different chemical evolution histories. In fact, star formation in dwarf galaxies 
are dominated by bursts and the particular star formation history of each object may be very different 
\citep[e.g.][]{lopezsanchezesteban10,karthicketal14}. In the diagram -- and considering the  dust correction commented above -- we can see that the Galactic disc models match quite well the distribution of the points corresponding to spiral galaxies. Chemical evolution models 
for local dwarf spheroidal galaxies (dSphs), as those by \citet{carigietal02}, are able to reproduce the range of C/O and O/H ratios observed in \hii\ regions belonging to dwarf irregular 
galaxies. This agreement is due to the similarity between the early bursts of star formation in dSphs and the recent bursts of star formation in dwarf irregulars 
\citep[see][their figure 9]{skillmanetal14}. 

In ionized nebulae, the C/O ratio can also be determined from the intensity of CELs of \ion{C}{2+} and \ion{O}{2+} in the UV. Based on that kind of lines, \citet{garnettetal95, garnettetal97},  \citet{kobulnickyetal97},  \citet{kobulnickyskillman98}, and \cite{izotovthuan99} determined the C/O ratios for a number of metal-poor galaxies. The same was done by \cite{garnettetal99} for a sample of {\hii} regions in the inner discs of spiral galaxies. Figure~\ref{cocomp} compares the C/O ratios in objects for which we have determinations based on both kinds of lines, CELs and RLs. The data based on RLs are from \citet[NGC~5253]{lopezsanchezetal07}, \citet[30~Dor]{apeimbert03} and  \citet[NGC~2363, NGC~5461, and VS~44]{estebanetal09}. The data based on CELs are from:  \citet[NGC~5253]{kobulnickyetal97}, \citet[30~Dor, and NGC~2363]{garnettetal95} and \citet[NGC~5461, and VS~44]{garnettetal99}. In  the cases of NGC~5461 and VS~44,  \citet{garnettetal99} 
give two values of the C/O ratio depending on two different reddening laws they assume in the UV. Figure~\ref{cocomp} shows that C/O ratios calculated with CELs and RLs are consistent within the uncertainties except in the case of NGC~2363.  The C/O ratio obtained by \cite{estebanetal09} from RLs for this object is substantially higher than that obtained by \cite{garnettetal95} from CELs perhaps because we observe a different zone of the object.

  \begin{figure}
   \centering
   \includegraphics[scale=0.42]{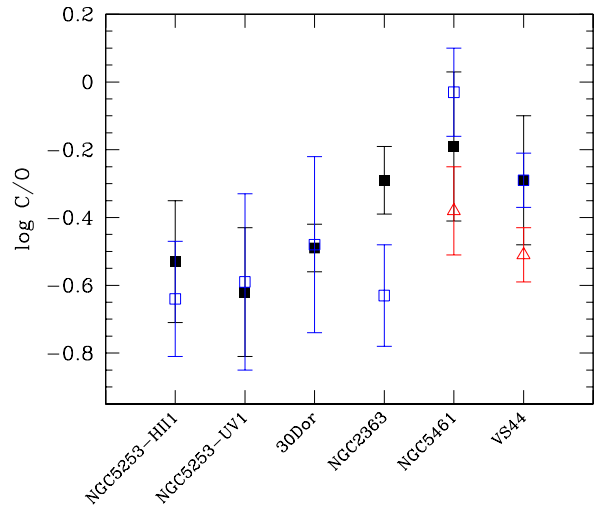} 
   \caption{C/O for extragalactic \hii\ regions determined from recombination lines (RLs, black filled squares) as well as from collisionally excited lines (CELs) in the UV (open colored symbols) for extragalactic {\hii} regions where these two determinations are available in the literature. 
   In the case of NGC~5461 and VS~44, the blue open squares and the red open triangles represent values obtained assuming two different reddening laws in the UV. 
   See \S\ref{covsoh} for references.  }
   \label{cocomp}
  \end{figure}    	 

Assuming that C/O(RLs) $\sim$ C/O(CELs) as Figure~\ref{cocomp} suggests, we have constructed Figure~\ref{covso4} including the same objects as in Figure~\ref{covso} but adding data for additional metal-poor star-forming dwarf galaxies obtained from UV CELs. These last data have been taken from  \citet{garnettetal97},  \citet{kobulnickyskillman98}, and  \citet{izotovthuan99}. The figure permits to explore the C/O {\it vs.}~O/H relation at lower metallicities. In the sake of consistency and for comparing O abundances determined using the same method, we have represented the O/H ratio determined from CELs and {\ts} = 0 for all the objects. For our opinion, this is the best solution because objects which C  abundance has been determined from UV CELs do not have estimate of their expected {\adfo} as well as of their corresponding {\ts} parameter. 
Now, in Figure~\ref{covso4}, the points that were included in Figure~\ref{covso} -- for 
which we determined O/H from RLs -- are displaced between 0.20 and 0.30 dex towards lower O abundances due to the aforementioned 
abundance discrepancy problem. The amount of such discrepancy for each object can be noted comparing the O abundance of the points having the same C/O ratios in figures~\ref{covso} and \ref{covso4}. 
From Figure~\ref{covso4}, it is now more evident that the objects show a clear trend in the sense of increasing the C/O ratio as the O abundance increases. 
This tendency has been interpreted as the combination of two 
processes: a) the time delay in the release of C by low- and intermediate-mass (LIM) stars with respect to the O production and b) metallicity-dependent yields of C in massive stars \citep{garnettetal99,henryetal00,carigi00,chiappinietal03}. The 
objects that show some displacement from that trend are indicated in Figure~\ref{covso4}: NGC~2363, UM~469,  and the two points corresponding to the NE and SW zones of I~Zw~18. 
However, UM~469 presents a large error and its  position should be even considered compatible with the general trend considering the uncertainties. 

  \begin{figure}
   \centering
   \includegraphics[scale=0.42]{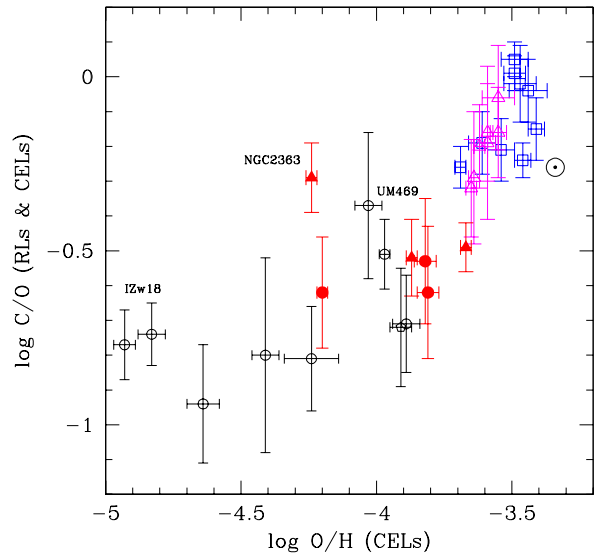} 
   \caption{C/O {\it vs.}~O/H for  \hii\ regions in the Milky Way and spiral, irregular and dwarf star-forming galaxies. The O/H ratios are obtained from CELs (considering {\ts} = 0) and the C/O ratios are determined from either CELs or RLs. Open (blue) squares represent data for Galactic \hii\ regions; open (magenta) triangles correspond to  {\hii} regions in the inner discs of external spiral galaxies; full (red) triangles indicate abundance ratios for {\hii} regions in irregular galaxies; full (red) circles correspond to dwarf star-forming galaxies (see references in the caption of Figure~\ref{covso}). Open (black) pentagons represent data for {\hii} regions in irregular galaxies obtained by \citet{garnettetal95}; open (black) circles correspond to data for dwarf star-forming galaxies from the 
   works by \citet{garnettetal97}, \citet{kobulnickyskillman98} and \citet{izotovthuan99}. The solar symbol represents the abundances of the Sun \citep{asplundetal09}. We recommend to add $\sim$0.1 dex to the log(O/H) values of {\hii} regions to correct for dust depletion and compare appropriately with solar abundances. No correction is needed for C/O ratios (see \S\ref{covsoh} for details). }
   \label{covso4}
  \end{figure}    	 

Figure~\ref{covsohalo3} further extends the C/O {\it vs.}~O/H diagram 
showing the behaviour of very metal-poor objects: stars of the Galactic halo \citep{fabbianetal09} and damped Ly$\alpha$ (DLA) systems \citep{cookeetal11}. For the Milky Way, it appears that the amount of C in halo stars is dominated by the contribution from massive stars \citep[e.g.][]{akermanetal04,fabbianetal09}. 
Figure~\ref{covsohalo3} shows that the position of metal-poor star-forming dwarf galaxies and halo stars is similar, suggesting the same origin for the 
production of the bulk of C in those galaxies. \citet{garnettetal95} came to the same conclusion from the comparison of the C/O ratio in dwarf galaxies with the predictions for massive star nucleosynthesis models of \citet{weaverwoosley93}.  The increase of the C/O toward the lowest metallicities may be the fingerprints of the contribution of Population III stars or the enhanced production of C induced by fast rotation in Population II stars \citep[e.g.][]{fabbianetal09}. It is interesting to note that the two points of I~Zw~18 -- that were slightly above the extrapolation
of the behaviour of other \hii\ regions with higher O/H ratios in Figure~\ref{covso4} -- match the general trend of the Galactic halo stars. From the data represented in Figure~\ref{covsohalo3}, a minimum of the C/O ratio seems to exist at log(O/H) $\sim$ $-$4.6. In any case, we must remind that the O abundances for the {\hii} regions represented in Figure~\ref{covsohalo3} have been determined making use of 
CELs and assuming {\ts} = 0. Values obtained from RLs would increase the O/H ratios about 0.2 to 0.3 dex in average.  

  \begin{figure}
   \centering
   \includegraphics[scale=0.42]{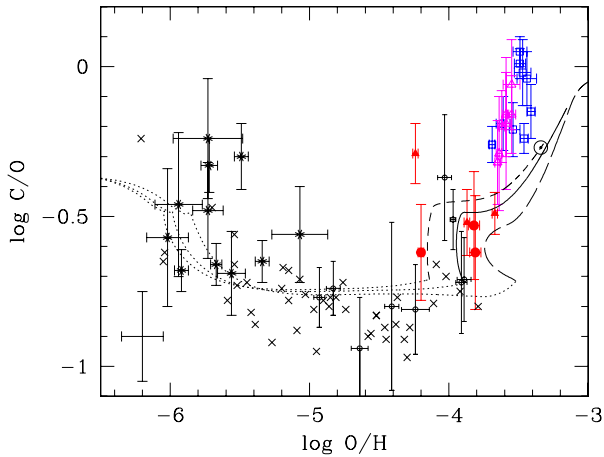} 
   \caption{The same as in Figure~\ref{covso4} but also including data for very metal-poor objects: crosses correspond to Galactic halo stars \citep{fabbianetal09} and asterisks with error bars to DLA systems \citep{cookeetal11}. The typical errorbar for halo stars data is at the bottom left corner of the panel. The lines show the predictions of chemical evolution models presented in Figure~\ref{covso}. We recommend to add $\sim$0.1 dex to the $log$(O/H) values of {\hii} regions to correct for dust depletion. No correction is needed for their C/O ratios (see \S\ref{covsoh} for details). }
   \label{covsohalo3}
  \end{figure}    	 

Figure~\ref{covsodisc} compares the position of the objects represented in Figure~\ref{covso} with the distribution of abundance ratios in 
nearby F and G dwarf stars of the Galactic disc \citep{bensbyfeltzing06}. In this figure, we have used the O/H ratios determined from RLs instead of CELs. This is because using CELs, 
the systematic offset between stellar and nebular abundances is between 0.3 and 0.5 dex, much larger than the 0.1 dex expected due to dust depletion. 
As we discussed in \S\ref{covsoh}, the C/O ratios seem to be not significantly affected by neither dust depletion nor the abundance discrepancy problem.  
\citet{bensbyfeltzing06} note that the C/O {\it vs.}~O/H trend is fairly similar to the well-known Fe/O {\it vs.}~O/H one, suggesting that LIM stars should be important contributors to the C enrichment at higher metallicities. This was also one of the main results of the chemical evolution models by \citet{carigietal05}. In Figure~\ref{covsodisc}, we can note that the segregation between {\hii} regions belonging to massive and low-mass galaxies is also reflected in the chemical properties of stars of the Galactic thin and thick discs. The locus defined by the stars of the thick disc coincides with that occupied by  
\hii\ regions in star-forming dwarf galaxies. In contrast, the locus of the thin disc stars is similar to that of \hii\ regions in the inner discs of the more massive spiral galaxies. 
The origin of the Galactic thick disc is still controversial. There are two main competing scenarios: a) it is a result of an ancient merger event between the Milky Way and another dwarf galaxy 
\citep[e.g.][]{quinnetal93, villaloboshelmi09} and b) it is the product of radial mixing of gas and stars from the thin disc  \citep[e.g.][]{schonrichbinney09}. In the light of the comparison provided in Figure~\ref{covsodisc}, the C/O {\it vs.}~O/H relation seems to favor the galaxy merger scenario for the origin of the Galactic thick disc, and that the stellar system 
absorbed by the Milky Way should be a dwarf galaxy. 

  \begin{figure}
   \centering
   \includegraphics[scale=0.42]{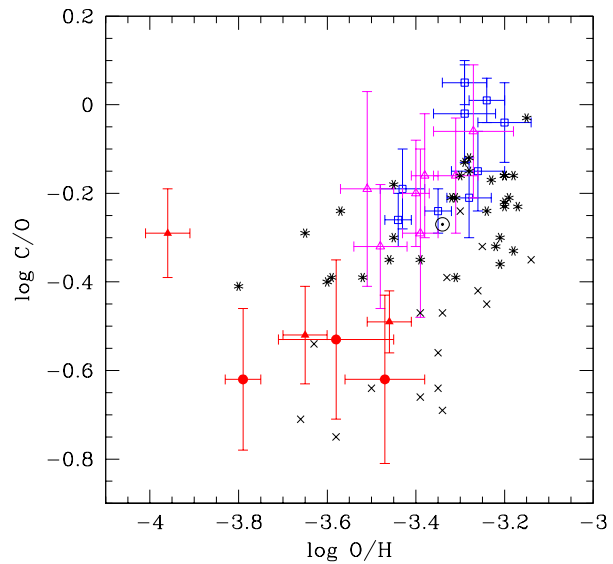} 
   \caption{The same as in Figure~\ref{covso} but also including data for F and G dwarf stars of the Galactic thin and thick discs (asterisks and crosses, respectively) obtained by \citet{bensbyfeltzing06}. 
   We recommend to add $\sim$0.1 dex to the log(O/H) values of {\hii} regions to correct for dust depletion. No correction is needed for their C/O ratios (see \S\ref{covsoh} for details). }
   \label{covsodisc}
  \end{figure}    	 
 
 \subsection{The C/O {\it vs.} N/O relation} \label{covno}
 
Using abundance values determined from UV and optical CELs, \citet{kobulnickyskillman98} found a possible correlation between C/O and N/O depending on the reddening law they assumed. The exploration of such relation is of much interest to compare the enrichment timescales of C and N. Figure~\ref{covno2} plots the C/O values of the 
objects included in Figure~\ref{covso4} with respect to their corresponding N/O ratios. This figure also includes  data for halo stars of the unmixed subsample of \citet{spiteetal05}, very metal-poor DLAs \citep{cookeetal11} 
and a carbon-enriched very metal-poor DLA reported by \citet{cookeetal12}. For \hii\ regions, 
the N/O ratios have been determined from the intensity of CELs and assuming {\ts} = 0, because there are no sufficiently bright pure RLs of N in the optical. The N/O ratios of the objects which C/O ratios have been derived from RLs -- those included in Figure~\ref{covso} -- have been taken from the same works as their C/O values  \citep[this work;][]{apeimbert03, estebanetal09, estebanetal13, garciarojasesteban07, lopezsanchezetal07}. For the {\hii} regions which C/O ratio was derived from UV CELs -- black open circles and pentagons -- their N/O ratios have been taken from the revision of optical emission-line flux ratios of \citet{navaetal06}, except in the case of SMC N88A for which we used the value obtained by \citet{testoretal03}. Contrary to what happens with C and O, N is expected not being a major constituent of dust, due to its inclusion in the highly stable gas form of N$_2$ \citep{gailsedlmayr86,jenkins14}. In the case of  Figure~\ref{covno2} we 
recommend to subtract $\sim$0.1 dex to the $log$(N/O) values of {\hii} regions to correct for dust depletion in O. Figure~\ref{covno2} shows that \hii\ region data are tightly correlated, with a slope close to 45$^\circ$. This trend is far more clear and includes more objects than that obtained by \citet{kobulnickyskillman98} and, moreover, it is independent of the reddening law assumed because it only involves optical emission lines for most of the objects. From the figure, we also see that the chemical evolution models for the Milky Way presented 
by \citet{estebanetal13} reproduce the position of \hii\ regions in the inner discs of spiral galaxies and the slope of the correlation. However, the models fail to reproduce the position of the halo 
stars, perhaps because of variations on the initial mass function, the star formation rate or even stochastic effects due to low numbers of massive stars polluting the gas where halo stars born \citep{carigihernandez08}. 

  \begin{figure}
   \centering
   \includegraphics[scale=0.42]{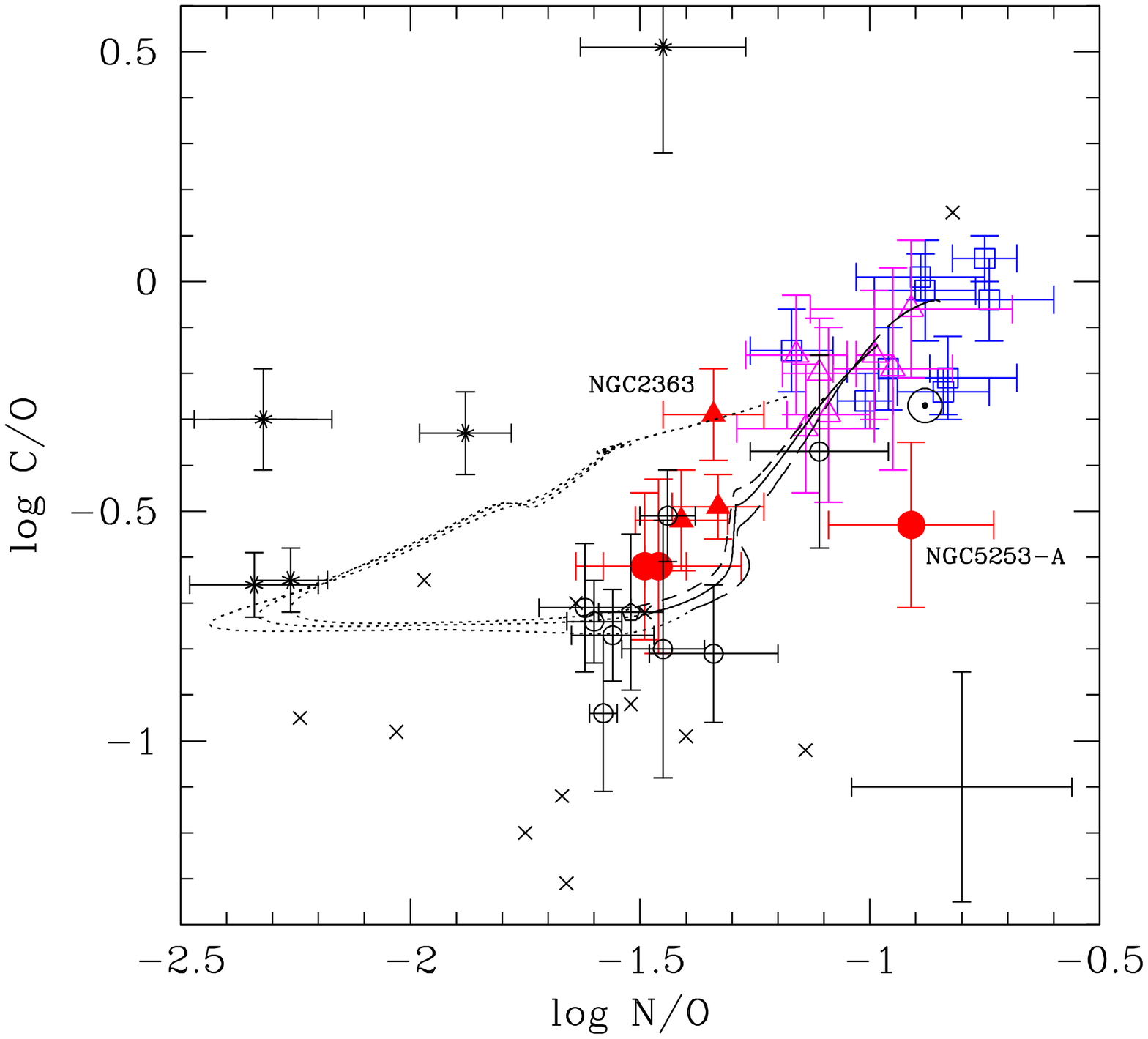} 
   \caption{C/O {\it vs.}~N/O for \hii\ regions in the Milky Way, spiral, irregular and star-forming dwarf galaxies as well as for halo stars \citep[][their unmixed subsample]{spiteetal05} and very metal poor DLAs \citep{cookeetal11, cookeetal12}. Symbols represent the same as in Figure~\ref{covso4} but now 
   we include data for a carbon enriched metal-poor DLA reported by \citet{cookeetal12} (the object at C/O = 0.51). In the case of {\hii} regions, the C/O ratios are obtained from the intensity of either RLs or CELs and the N/O values have been derived from CELs (see text for details and references). The typical errorbar for halo stars data is at the bottom right corner of the panel. The lines show the predictions of chemical evolution models presented in Figure~\ref{covso}. We recommend to subtract $\sim$0.1 dex to the log(N/O) values of {\hii} regions to correct for dust depletion in O (see \S\ref{covno} for details). }
   \label{covno2}
  \end{figure}    	 

  \begin{figure}
   \centering
   \includegraphics[scale=0.42]{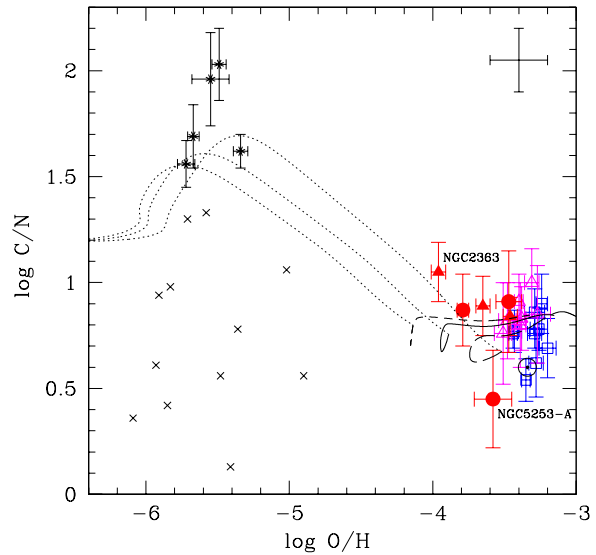} 
   \caption{C/N {\it vs.}~O/H for Galactic and extragalactic \hii\ regions, halo stars and very metal poor DLAs. Symbols represent the same as in Figure~\ref{covno2}. 
   In the case of {\hii} regions, C and O abundances are determined from RLs and N ones from CELs and assuming \ts\ $>$ 0 (see \S\ref{covno} for details). The typical errorbar for halo stars data is at the upper right corner of the panel. The lines show the predictions of chemical evolution models presented in Figure~\ref{covso}. We recommend to add $\sim$0.1 dex to the values of both log(C/N) and log(O/H) of {\hii} regions to correct for dust depletion (see \S\ref{covno} for details).}
   \label{cnvso2}
  \end{figure}    	 
 
The apparent coupling between the C and N enrichment we see from \hii\ region data is further suggested in Figure~\ref{cnvso2}, that plots the C/N {\it vs.}~O/H relation for \hii\ regions which C/O ratios derived from RLs as well as halo stars \citep[][unmixed subsample]{spiteetal05}  and very metal-poor DLAs \citep{cookeetal11, cookeetal12}. In nebulae, 
for calculating the C/N ratio we have assumed the N abundances determined from CELs corrected for the presence of temperature fluctuations using the values of \ts\ parameter estimated for each object. This is performed to correct for the abundance discrepancy problem assuming that it is related 
to temperature variations in the nebulae \citep[e.g.][]{garciarojasesteban07}. On the other hand, the quantities represented in both axes  of Figure~\ref{cnvso2} are expected to be affected by some amount of 
dust depletion. As we discussed in \S\ref{covsoh}, we should add $\sim$0.1 dex to both log(C/N) and log(O/H) of {\hii} regions in order to correct for dust depletion.  

From Figure~\ref{cnvso2} we can note that, for \hii\ region data, C/N {\it vs.}~O/H does not show a clear correlation, indicating a rather constant value of C/N in the range of metallicities that those objects occupy. Other authors have explored this relation using abundance data determined from UV and optical CELs, finding either a slight tendency for lower C/N ratios at lower metallicity \citep{kobulnickyskillman98} or a lack of correlation \citep{garnettetal99} as in our case. 
The chemical evolution models for the Milky Way of \citet{estebanetal13} are able to reproduce the rather constant C/N ratio since the formation of the Galactic disc. 

The apparent coupling between the C and N enrichment that we observe from \hii\ region data at high metallicities is due to the cumulative-temporal contribution of the C and N yields 
by stars of different masses. The nucleosynthesis processes of both elements are different, but their enrichment timescales result to be rather similar, due to the metallicity-dependent yields 
and the lifetimes of stars that contribute to the C and N increase.  Following 
\citet{carigietal05}, at high metallicities, log(O/H) $\geq$ $-$3.5, the main contributors to C and N should be massive stars. However, at intermediate metallicities,  $-$3.9 $\leq$ log(O/H) $\leq$ $-$3.5, the main contributors to C are low metallicity low-mass stars, but for N the main producers are massive stars and intermediate-mass stars of intermediate and low metallicity, respectively. Figure~\ref{cnvso2} shows a very large 
dispersion of the C/N ratio for halo stars, indicating that at very low metallicities the apparent coupling between N and C enrichments breaks. Here stochastic effects, bursting star formation  or local variations on the enrichment process in the first stages of the galaxy formation could affect the abundance ratio of these elements across the halo. 

The peculiar position of the star-forming knot of NGC~5253 labelled as A by \citet{lopezsanchezetal07} in Figures~\ref{covno2} and \ref{cnvso2} is due to the effect of localized pollution of  material processed by the CNO cycle in massive Wolf-Rayet stars, as it has been confirmed by the results of different authors \citep{welch70, walshroy87, walshroy89, kobulnickyetal97, lopezsanchezetal07,monreal-iberoetal10}. This knot shows strong enhancement of N -- and perhaps He -- while the abundances of C and O remain quite constant across the object \citep[e.g.][]{lopezsanchezetal07}. This is the only star-forming dwarf galaxy where chemical pollution by massive Wolf-Rayet stars has been unambiguously detected. 
On the other hand, as we commented in \S\ref{covsoh}, NGC~2363 is in a peculiar position in Figures~\ref{covso} and \ref{covso4} due 
to its high C/O ratio. In principle, following the arguments of  \citet{garnettetal95} to explain the excess C/O ratios they found for I~Zw~18, this fact may be interpreted as the effect of delayed ejection of fresh C from LIM stars due to a previous star-formation event in the host irregular galaxy. If this is the case, this object should also 
show some increase of the N/O ratio with respect to its metallicity due to the contribution of N produced by LIM stars. The position of NGC~2363 in Figure~\ref{covno2} 
seems to support this hypothesis because it follows the general C/O {\it vs.}~N/O correlation and therefore the excess C/O seems to be associated with a somewhat higher N/O ratio.  However, the position of NGC~2363 in Figure~\ref{cnvso2} seems to be slighly above the rest of the objects, and the presence of a small excess of C in the observed area of this 
object can not be ruled out.

\section{Conclusions} \label{conclusions}
 
 We present deep echelle spectrophotometry of the brightest emission-line knots of the star-forming galaxies He~2$-$10, Mkn~1271, NGC~3125, NGC~5408, POX~4, SDSS~J1253$-$0312, Tol~1457$-$262, Tol~1924$-$416, and the {\hii} region Hubble V in the Local Group dwarf irregular galaxy NGC~6822. The spectra cover from 3550 to 10 400~{\AA} and have been taken with the Ultraviolet-Visual Echelle Spectrograph attached to the UT2 of the Very Large Telescope at Cerro Paranal Observatory. We have derived consistent and precise values of the physical conditions for each object making use of several emission line-intensity ratios as well as abundances for several ionic species from the intensity of 
 collisionally excited lines (CELs). We derive the ionic abundances of C$^{2+}$ and/or O$^{2+}$ from faint pure recombination lines (RLs) 
 in several of the objects, permitting to derive their C/H and C/O ratios. For some objects, we have estimated the {\ts} parameter from the analysis of the intensity ratios of {\hei} lines as well as assuming that the {\adfo} is produced by temperature fluctuations, finding that both methods give in general consistent results. We find also that giant {\hii} regions in dwarf star-forming galaxies present the highest {\adfo}  and {\ts} values among the inventory of determinations available for {\hii} regions.
 
 We have explored the chemical evolution in low-metallicity objects analysing the C/O {\it vs.}~O/H relation and comparing with model results for the Milky Way. We find that {\hii} regions in 
 star-forming dwarf galaxies tend to show lower values of the C/O ratio and somewhat lower O/H than {\hii} regions belonging to spiral galaxies, indicating their different chemical evolution histories. The comparison with C/O ratios in other objects indicate that the position of star-forming dwarf galaxies is similar to that of Galactic halo stars, suggesting the same origin for the bulk of C in those galaxies. We have also found that the dichotomy between the C/O ratios observed in stars of the Galactic thick and thin discs coincides with the systematical 
 differences between the C/O ratios in {\hii} regions belonging to spiral or dwarf star-forming galaxies. This fact supports the merging scenario for the origin of the Galactic thick disc. 
 Finally, we explore the C/O {\it vs.}~N/O and C/N {\it vs.}~O/H relations, finding that -- at the usual metallicities determined for {\hii} regions -- there is an apparent coupling between C and N enrichment that may be due to the cumulative-temporal contribution of the C and N by stars of different masses. For very low-metallicity objects -- as Galactic halo stars-- such coupling breaks. This may be due to the bursting star-formation mode, stochastic effects, and/or local pollution in the first stages of the galaxy formation.

\section*{Acknowledgments}
We are very grateful to the referee, Grazyna Stasi\'nska, for her very valuable and constructive comments and suggestions that have contributed to improve the final version of the paper. This work has been funded by the Spanish Ministerio de Econom\'\i a y Competitividad (MINECO) under project AYA2011-22614. L. Carigi and M. Peimbert are grateful to the financial support provided by CONACyT 
 of Mexico (grant 129753). L. Carigi also thanks financial support from project AYA2010-16717 funded by MINECO. A. Mesa-Delgado acknowledges support from a Basal-CATA (PFB-06/2007) grant and the FONDECYT project 3140383. This research has made use of the NASA/IPAC Extragalactic Database (NED) which is operated by the Jet Propulsion Laboratory, California Institute of Technology, under contract with
the National Aeronautics and Space Administration. This research has made use of the SIMBAD database,
operated at CDS, Strasbourg, France. 


\label{lastpage}

\end{document}

%% file: table_lines_1.tex
   \begin{table*}
  \centering
   \begin{minipage}{210mm}
     \caption{Dereddened line intensity ratios with respect to $I$(\hb) = 100 of HV (NGC~6822), NGC~5408, and Tol~1924$-$416.}
     \label{lines1}
\begin{tabular}{cccccccccc}
     \hline                                              
        & & & & \multicolumn{2}{c}{HV (NGC~6822)} & \multicolumn{2}{c}{NGC~5408} & \multicolumn{2}{c}{Tol~1924$-$416} \\  
       $\lambda_0$ & & & & $\lambda$ & & $\lambda$ & & $\lambda$ & \\
       (\AA) & Ion & ID& $f$($\lambda$) & (\AA) & $I$($\lambda$) & (\AA) & $I$($\lambda$) & (\AA) & $I$($\lambda$) \\ 
     \hline        
3187.84     &    {\hei}    &       3      &     0.308    &    3187.16   &     0.72     $\pm$    0.07     &    3193.11   &     1.13     $\pm$    0.09       &     3217.88     &       1.7     $\pm$    0.2     \\
3342.85     &   {\fcliii}  &      1F      &     0.304    &      $-$     &               $-$              &    3348.17   &     0.29     $\pm$    0.05       &       $-$       &                $-$             \\
3447.59     &    {\hei}    &       7      &     0.295    &    3446.92   &     0.30     $\pm$    0.04     &    3453.30   &     0.18     $\pm$    0.04       &     3480.17     &       0.4     $\pm$    0.1     \\
3450.39     &   {\hei}  &          &     0.295    &      $-$     &               $-$              &    3457.80   &     0.06:       &       $-$       &                $-$             \\
3530.50     &    {\hei}    &      36      &     0.286    &      $-$     &               $-$              &    3536.38   &     0.15     $\pm$    0.04       &       $-$       &                $-$             \\
\end{tabular}
\end{minipage}
\end{table*}
\setcounter{table}{1}
   \begin{table*}
  \centering
   \begin{minipage}{210mm}
     \caption{\it continued}
     \label{lines1}
    \begin{tabular}{cccccccccc}
     \hline                                              
        & & & & \multicolumn{2}{c}{HV (NGC~6822)} & \multicolumn{2}{c}{NGC~5408} & \multicolumn{2}{c}{Tol~1924$-$416} \\  
       $\lambda_0$ & & & & $\lambda$ & & $\lambda$ & & $\lambda$ & \\
       (\AA) & Ion & ID& $f$($\lambda$) & (\AA) & $I$($\lambda$) & (\AA) & $I$($\lambda$) & (\AA) & $I$($\lambda$) \\ 
     \hline 
4209.15     &   {\ffeiii}  &              &     0.157    &    4208.45   &     0.027    $\pm$    0.007    &      $-$     &               $-$                &       $-$       &                $-$             \\
4227.91     &   {\fniqiii}  &              &     0.153    &      $-$     &               $-$              &    4234.77   &     0.09     $\pm$    0.01       &       $-$       &                $-$             \\
4243.97     &   {\ffeii}   &      21F     &     0.149    &      $-$     &               $-$              &    4251.08   &     0.027    $\pm$    0.007      &     4284.05     &      0.07     $\pm$    0.02    \\
4267.15     &    {\cii}    &       6      &     0.144    &    4266.36   &     0.046    $\pm$    0.008    &    4274.18   &     0.020    $\pm$    0.006      &       $-$       &                $-$             \\
4276.83     &   {\ffeii}   &      21F     &     0.142    &      $-$     &               $-$              &    4283.85   &     0.039    $\pm$    0.008      &     4317.37     &      0.12     $\pm$    0.03    \\
\end{tabular}
\end{minipage}
\end{table*}
\setcounter{table}{1}
   \begin{table*}
  \centering
   \begin{minipage}{210mm}
     \caption{\it continued}
     \label{lines1}
    \begin{tabular}{cccccccccc}
     \hline                                              
        & & & & \multicolumn{2}{c}{HV (NGC~6822)} & \multicolumn{2}{c}{NGC~5408} & \multicolumn{2}{c}{Tol~1924$-$416} \\  
       $\lambda_0$ & & & & $\lambda$ & & $\lambda$ & & $\lambda$ & \\
       (\AA) & Ion & ID& $f$($\lambda$) & (\AA) & $I$($\lambda$) & (\AA) & $I$($\lambda$) & (\AA) & $I$($\lambda$) \\ 
     \hline 
4985.9      &   {\ffeiii}  &      2F      &    -0.031    &    4984.91   &     0.17     $\pm$    0.01     &    4994.25   &     0.31     $\pm$    0.02       &     5033.00     &      1.12     $\pm$    0.08    \\
5006.84     &   {\foiii}   &      1F      &    -0.036    &    5005.96   &     541    $\pm$    22     &    5015.30   &      544     $\pm$     19        &     5054.14     &       608     $\pm$     21     \\
5015.68     &    {\hei}    &       4      &    -0.038    &    5014.78   &     2.21     $\pm$    0.09     &    5024.11   &     1.98     $\pm$    0.08       &     5063.05     &       2.0     $\pm$    0.1     \\
5041.03     &   {\silii}   &       5      &    -0.044    &    5040.02   &     0.09:     &    5049.48   &     0.11     $\pm$    0.02       &       $-$       &                $-$             \\
5055.98     &   {\silii}   &       5      &    -0.048    &      $-$     &               $-$              &    5064.52   &     0.05     $\pm$    0.01       &       $-$       &                $-$             \\
\end{tabular}
\end{minipage}
\end{table*}
\setcounter{table}{1}
   \begin{table*}
  \centering
   \begin{minipage}{210mm}
     \caption{\it continued}
     \label{lines1}
    \begin{tabular}{cccccccccc}
     \hline                                              
        & & & & \multicolumn{2}{c}{HV (NGC~6822)} & \multicolumn{2}{c}{NGC~5408} & \multicolumn{2}{c}{Tol~1924$-$416} \\  
       $\lambda_0$ & & & & $\lambda$ & & $\lambda$ & & $\lambda$ & \\
       (\AA) & Ion & ID& $f$($\lambda$) & (\AA) & $I$($\lambda$) & (\AA) & $I$($\lambda$) & (\AA) & $I$($\lambda$) \\ 
     \hline 
8264.28     &     {\hi}    &      P35     &    -0.517    &    8262.87   &     0.069    $\pm$    0.008    &      $-$     &               $-$                &       $-$       &                $-$             \\
8267.94     &     {\hi}    &      P34     &    -0.518    &    8266.43   &     0.061    $\pm$    0.008    &      $-$     &               $-$                &       $-$       &                $-$             \\
8271.93     &     {\hi}    &      P33     &    -0.518    &    8270.48   &     0.09     $\pm$    0.01     &      $-$     &               $-$                &       $-$       &                $-$             \\
8276.31     &     {\hi}    &      P32     &    -0.518    &    8274.63   &     0.037    $\pm$    0.006    &    8290.50   &     0.09     $\pm$    0.01       &       $-$       &                $-$             \\
8286.43     &     {\hi}    &      P30     &    -0.519    &    8284.94   &     0.09     $\pm$    0.01     &      $-$     &               $-$                &       $-$       &                $-$             \\
\hline                                              
    \end{tabular}
    \begin{description}
      \item[$^{\rm a}$] Blend of {\foii} $\lambda$7318.39 and $\lambda$7319.99 lines.  
      \item[$^{\rm b}$] Blend of {\foii} $\lambda$7329.66 and $\lambda$7330.73 lines.                   
    \end{description}
   \end{minipage}
   \end{table*}

%% file: table_lines_2.tex
   \begin{table*}
  \centering
   \begin{minipage}{210mm}
     \caption{Dereddened line intensity ratios with respect to $I$(\hb) = 100 of Mkn~3125, Mkn~1271, and POX~4.}
     \label{lines2}
    \begin{tabular}{cccccccccc}
     \hline                                              
        & & & & \multicolumn{2}{c}{NGC~3125} & \multicolumn{2}{c}{Mkn~1271} & \multicolumn{2}{c}{POX~4} \\  
       $\lambda_0$ & & & & $\lambda$ & & $\lambda$ & & $\lambda$ & \\
       (\AA) & Ion & ID& $f$($\lambda$) & (\AA) & $I$($\lambda$) & (\AA) & $I$($\lambda$) & (\AA) & $I$($\lambda$) \\ 
     \hline        
3187.84      &    {\hei}     &       3        &    0.308     &   3199.36   &     1.0   $\pm$   0.1     &   3198.69   &     1.0   $\pm$   0.2     &   3225.32   &     1.3   $\pm$   0.2    \\
3447.59      &    {\hei}     &       7        &    0.295     &   3460.10   &     0.3   $\pm$   0.1     &     $-$     &            $-$            &     $-$     &            $-$           \\
3587.28      &    {\hei}     &       3        &    0.278     &   3600.45   &     0.4   $\pm$   0.1     &     $-$     &            $-$            &     $-$     &            $-$           \\
3613.64      &    {\hei}     &       6        &    0.275     &   3626.81   &     0.3   $\pm$   0.1     &     $-$     &            $-$            &     $-$     &            $-$           \\
3634.25      &    {\hei}     &       28       &    0.272     &   3647.34   &     0.3   $\pm$   0.1     &     $-$     &            $-$            &   3677.03   &     0.3   $\pm$   0.1    \\
\end{tabular}
\end{minipage}
\end{table*}
\setcounter{table}{2}
   \begin{table*}
  \centering
   \begin{minipage}{210mm}
     \caption{\it continued}
     \label{lines2}
    \begin{tabular}{cccccccccc}
     \hline                                              
        & & & & \multicolumn{2}{c}{NGC~3125} & \multicolumn{2}{c}{Mkn~1271} & \multicolumn{2}{c}{POX~4} \\  
       $\lambda_0$ & & & & $\lambda$ & & $\lambda$ & & $\lambda$ & \\
       (\AA) & Ion & ID& $f$($\lambda$) & (\AA) & $I$($\lambda$) & (\AA) & $I$($\lambda$) & (\AA) & $I$($\lambda$) \\ 
     \hline 
4708.04      &   ?   &                &    0.038     &   4725.11   &    0.03   $\pm$    0.01      &     $-$     &            $-$            &     $-$     &            $-$           \\
4711.37      &   {\fariv}    &       1F       &    0.037     &   4728.48   &    0.37   $\pm$  0.04     &   4727.54   &    1.03   $\pm$  0.09     &   4766.80   &    1.10   $\pm$  0.07    \\
4713.14      &    {\hei}     &       12       &    0.036     &   4730.27   &    0.48   $\pm$  0.04     &   4729.30   &    0.54   $\pm$  0.07     &   4768.51   &    0.56   $\pm$  0.05    \\
4740.16      &   {\fariv}    &       1F       &    0.030     &   4757.40   &    0.27   $\pm$  0.03     &   4756.45   &    0.76   $\pm$  0.08     &   4796.02   &    0.98   $\pm$  0.07    \\
4754.83      &   {\ffeiii}   &       3F       &    0.026     &   4771.95   &    0.11   $\pm$  0.02     &     $-$     &            $-$            &     $-$     &            $-$           \\
\end{tabular}
\end{minipage}
\end{table*}
\setcounter{table}{2}
   \begin{table*}
  \centering
   \begin{minipage}{210mm}
     \caption{\it continued}
     \label{lines2}
    \begin{tabular}{cccccccccc}
     \hline                                              
        & & & & \multicolumn{2}{c}{NGC~3125} & \multicolumn{2}{c}{Mkn~1271} & \multicolumn{2}{c}{POX~4} \\  
       $\lambda_0$ & & & & $\lambda$ & & $\lambda$ & & $\lambda$ & \\
       (\AA) & Ion & ID& $f$($\lambda$) & (\AA) & $I$($\lambda$) & (\AA) & $I$($\lambda$) & (\AA) & $I$($\lambda$) \\ 
     \hline 
9545.97   & {\hi}     &  P8   &  -0.619&  9581.4&    1.3 $\pm$ 0.1&  $-$   &     $-$    &  9658.32&   3.4 $\pm$  0.3\\
10049.40  & {\hi}     &  P7   &  -0.650&  $-$   &       $-$     &  $-$   &     $-$    &  10167.5&   5.6 $\pm$  0.5\\
\multicolumn{4}{l}{$c$(H$\beta$)} & \multicolumn{2}{c}{0.45 $\pm$ 0.06} & \multicolumn{2}{c}{0.10 $\pm$ 0.06} & \multicolumn{2}{c}{0.20 $\pm$ 0.05} \\
\multicolumn{4}{l}{$F$(H$\beta$) (10$^{-13}$ erg cm$^{-2}$ s$^{-1}$)} & \multicolumn{2}{c}{1.78 $\pm$ 0.07} & \multicolumn{2}{c}{1.43 $\pm$ 0.06} & \multicolumn{2}{c}{2.6 $\pm$ 0.1} \\
\multicolumn{4}{l}{$-W$(H$\beta$) (\AA)} & \multicolumn{2}{c}{78} & \multicolumn{2}{c}{88} & \multicolumn{2}{c}{148} \\
\multicolumn{4}{l}{$W_{\rm abs}$ (\AA)} & \multicolumn{2}{c}{0.5} & \multicolumn{2}{c}{1.2} & \multicolumn{2}{c}{1.5}  \\ 
    \hline                                              
    \end{tabular}
    \begin{description}
      \item[$^{\rm a}$] Blend of {\foii} $\lambda$7318.39 and $\lambda$7319.99 lines.  
      \item[$^{\rm b}$] Blend of {\foii} $\lambda$7329.66 and $\lambda$7330.73 lines.                   
    \end{description}
   \end{minipage}
   \end{table*}

%% file: table_lines_3.tex
   \begin{table*}
  \centering
   \begin{minipage}{210mm}
     \caption{Dereddened line intensity ratios with respect to $I$(\hb) = 100 of SDSS~J1253$-$0312, Tol~1457$-$262, and He~2$-$10.}
     \label{lines3}
    \begin{tabular}{cccccccccc}
     \hline                                              
        & & & & \multicolumn{2}{c}{SDSS~J1253$-$0312} & \multicolumn{2}{c}{Tol~1457$-$262} & \multicolumn{2}{c}{He~2$-$10} \\  
       $\lambda_0$ & & & & $\lambda$ & & $\lambda$ & & $\lambda$ & \\
       (\AA) & Ion & ID& $f$($\lambda$) & (\AA) & $I$($\lambda$) & (\AA) & $I$($\lambda$) & (\AA) & $I$($\lambda$) \\ 
     \hline        
3187.84   &    {\hei}     &      3      &  0.308   & 3260.20 &  5.4  $\pm$  0.4   & 3241.72 &  2.1  $\pm$  0.4   & 3197.1 &  0.7  $\pm$   0.1     \\
3671.48   &     {\hi}     &     H24     &  0.266   &   $-$   &        $-$         &   $-$   &        $-$         & 3682.3 &  0.30  $\pm$   0.09     \\
3673.76   &     {\hi}     &     H23     &  0.265   &   $-$   &        $-$         &   $-$   &        $-$         & 3684.3 &  0.4  $\pm$   0.1     \\
3676.37   &     {\hi}     &     H22     &  0.265   & 3759.86 &  0.4  $\pm$  0.1   &   $-$   &        $-$         & 3686.9 &  0.5  $\pm$   0.1     \\
3679.36   &     {\hi}     &     H21     &  0.265   & 3762.91 &  0.4  $\pm$  0.1   &   $-$   &        $-$         & 3690.0 &  0.5  $\pm$   0.1     \\
\end{tabular}
\end{minipage}
\end{table*}
\setcounter{table}{3}
   \begin{table*}
  \centering
   \begin{minipage}{210mm}
     \caption{\it continued}
     \label{lines3}
    \begin{tabular}{cccccccccc}
     \hline                                              
        & & & & \multicolumn{2}{c}{SDSS~J1253$-$0312} & \multicolumn{2}{c}{Tol~1457$-$262} & \multicolumn{2}{c}{He~2$-$10} \\  
       $\lambda_0$ & & & & $\lambda$ & & $\lambda$ & & $\lambda$ & \\
       (\AA) & Ion & ID& $f$($\lambda$) & (\AA) & $I$($\lambda$) & (\AA) & $I$($\lambda$) & (\AA) & $I$($\lambda$) \\ 
     \hline 
5517.71   &   {\fcliii}   &     1F      &  -0.154  & 5643.20 & 0.22  $\pm$  0.06  & 5611.26 &  0.5  $\pm$  0.1   &  $-$   &         $-$             \\
5537.88   &   {\fcliii}   &     1F      &  -0.158  & 5663.83 & 0.18  $\pm$  0.05  & 5631.95 &  0.5  $\pm$  0.1   &  $-$   &         $-$             \\
5875.64   &    {\hei}     &     11      &  -0.215  & 6009.16 & 11.6  $\pm$  0.6   & 5975.12 & 11.4  $\pm$  0.5   &  $-$   &         $-$             \\
6300.30   &    {\foi}     &     1F      &  -0.282  & 6443.52 &  2.3  $\pm$  0.2   & 6406.79 &  4.5  $\pm$  0.3   & 6318.6 &  1.9  $\pm$   0.1     \\
6312.10   &   {\fsiii}    &     3F      &  -0.283  & 6455.59 &  1.6  $\pm$  0.1   & 6418.97 &  1.6  $\pm$  0.2   & 6330.3 &  0.94  $\pm$   0.07     \\
\hline                                              
    \end{tabular}
    \begin{description}
       \item[$^{\rm a}$] Line intensity severily affected by telluric absorption features. 
       \item[$^{\rm b}$] Blend of {\foii} $\lambda$7318.39 and $\lambda$7319.99 lines.  
      \item[$^{\rm c}$] Blend of {\foii} $\lambda$7329.66 and $\lambda$7330.73 lines.                   
    \end{description}
   \end{minipage}
   \end{table*}